\documentclass[aps,prb,twocolumn,,superscriptaddress,showpacs]{revtex4-1}
\usepackage[dvipdfmx]{graphicx,color}
\usepackage{amsmath,bm}
\newcommand{\beq}{\begin{equation}} \newcommand{\eeq}{\end{equation}}
 
\begin{document}
\title{Distribution of eigenstate populations and dissipative beating dynamics \\ in uniaxial single-spin magnets}
\author{Takuya Hatomura}
\email[]{hatomura@spin.phys.s.u-tokyo.ac.jp}
\affiliation{Department of Physics, Graduate School of Science, The University of Tokyo, 7-3-1 Hongo, Bunkyo, Tokyo, Japan}
\author{Bernard Barbara}
\affiliation{Institut N\'eel, CNRS and Universit\'e Grenoble-Alpes, B.P. 166 38042, Cedex 166, Grenoble, France}
\author{Seiji Miyashita}
\affiliation{Department of Physics, Graduate School of Science, The University of Tokyo, 7-3-1 Hongo, Bunkyo, Tokyo, Japan}
\affiliation{CREST, JST, 4-1-8 Honcho Kawaguchi, Saitama, Japan}
\date{\today}
%
%
\begin{abstract}
Numerical simulations of magnetization reversal of a quantum uniaxial magnet under a swept magnetic field [Hatomura, \textit{et al}., \textit{Quantum Stoner-Wohlfarth Model}, Phys. Rev. Lett. \textbf{116}, 037203 (2016)] are extended. 
In particular, how the ``wave packet" describing the time-evolution of the system is scattered in the successive avoided level crossings is investigated from the viewpoint of the distribution of the eigenstate populations. 
It is found that the peak of the distribution as a function of the magnetic field does not depend on spin-size $S$, which indicates that the delay of magnetization reversal due to the finite sweeping rate is the same in both the quantum and classical cases. 
The peculiar synchronized oscillations of all the spin components result in the beating of the spin-length. 
Here, dissipative effects on this beating are studied by making use of the generalized Lindblad-type master equation. 
The corresponding experimental situations are also discussed in order to find conditions for experimental observations. 
\end{abstract}
%
%
\maketitle
%
%
\section{\label{intro}Introduction}
Quantum spin systems are usually realized in experiments by diluted ensembles of single-domain ferromagnetic nano-particles~\cite{Neel,PhysRevLett.79.4014}, single-molecule magnets~\cite{SGCN,GSV}, or single-spins magnets~\cite{PhysRevLett.87.057203,PhysRevLett.91.257204}. 
These systems are generally with relatively large spins $S$ (collective spins of nano-particles and single-molecule magnets, or large rare-earth angular momentum of single-spins magnets). 
Except for the nano-particles case, this ensures the total absence of size- and therefore spin-distribution, and as well, slow enough quantum dynamics to be measurable. 
Due to small interactions, they can be regarded as almost isolated, even at low temperatures. 
This is particularly true with single-spin magnets which can be diluted at will. 
Various quantum effects have been reported in both single-molecule and single-spin magnets. 
One of the significant quantum effects is the stepwise magnetization process in hysteresis of blocked magnetization at low temperatures, in which quantum tunneling between the states with magnetizations in opposite directions plays an important role~\cite{TLBGSB,B}. 
This phenomenon is understood as a quantum hybridization of discrete opposite magnetization energy levels forming an avoided level crossing. 
The scattering rate at the avoided crossing point is characterized by the Landau-Zener formula~\cite{L,Z,M,S}, which has been theoretically applied to quantum nano-magnets~\cite{doi:10.1143/JPSJ.64.3207,doi:10.1143/JPSJ.65.2734,PhysRevB.56.11761} and confirmed in experiments~\cite{WS,UMK}. 
Quantum effects at each avoided crossing point have been intensively studied both theoretically and experimentally during the last decades~\cite{Owerre2015}. 

Uniform magnetization reversal in a single ferromagnetic domain associated with the classical metastable state collapse was studied by Stoner and Wohlfarth~\cite{SW}. 
Because of the uniaxial anisotropy, the magnetization opposite to the direction of the magnetic field is metastable until the magnetic field reaches a certain value. 
At the end of metastability, which is called the Stoner-Wohlfarth point, magnetization exhibits a jump to the stable direction. 
This point draws an astroid known as the Stoner-Wohlfarth astroid in the longitudinal and transverse fields' plane. 

In the previous study~\cite{HBM}, we considered this metastable-to-stable transition in the quantum case, where quantum tunneling at each anti-level crossing plays an important role. 
Besides the well-known stepwise hysteresis mentioned above, we discovered the spinodal-like critical behavior of the energy gaps in the successive avoided crossing points. 
This result was obtained by a study of the magnetization reversal resulting from the successive Landau-Zener scatterings at avoided crossing points. 
These scatterings take place along the continuation of the ground state energy level for $H_z>0$ to the $H_z<0$ until the Stoner-Wohlfarth point, which we called the metastable branch. 
The spinodal criticality appears along this metastable branch, or more precisely around the Stoner-Wohlfarth point, at which the metastable state in the corresponding classical model collapses into the stable state. 
Furthermore, a characteristic recursive beating of the magnetization takes place during its precession beyond the Stoner-Wohlfarth point. 
This beating results from the synchronized oscillations of all the components of magnetization and leads to the recursive oscillation of the spin-length, $s_f\equiv\langle s_x\rangle^2+\langle s_y\rangle^2+\langle s_z\rangle^2$, which we called the spin-fidelity. 

In the present paper, we study the nature of the transition from the quantum mechanical behavior to the classical one. 
In particular, we investigate how the distribution of the eigenstate populations for a given magnetic field $H_z$ changes as a function of $S$ and discuss the corresponding classical deterministic state. 
It is found that the distributions of the scattered populations beyond the Stoner-Wohlfarth point follow a universal scaling law independent of $S$. 
This indicates that the same distribution holds even in the classical case. 
As to the beating of magnetization, we study the dissipative effects by making use of the generalized Lindblad-type equation~\cite{NC,BP,W,STM}. 
The results enable us to provide the conditions to observe the beating phenomenon in experiments. 

This paper is constructed as follows. 
In Sec.~\ref{QSWmodel}, the model is explained. 
In particular, the notations of the quantum Stoner-Wohlfarth model and the conventional spin Hamiltonian are explained in detail. 
The distribution of the eigenstate populations beyond the Stoner-Wohlfarth point and its classical limit are studied 
in Sec.~\ref{distribution}. 
Section~\ref{dissipative} is devoted to the dissipative dynamics of the quantum Stoner-Wohlfarth model. 
We give summary and discussions in Sec.~\ref{sum}.

%
%
\section{\label{QSWmodel}Quantum Stoner-Wohlfarth model}
\subsection{Uniaxial single-spin magnets}
The spin Hamiltonian, which is used for the study of uniaxial quantum magnets, such as the molecular magnet $\mathrm{Mn}_{12}$, is generally written as
\beq
\tilde{\mathcal{H}}=-\tilde{D}S_z^2-\tilde{H}_xS_x-\tilde{H}_zS_z,
\label{Eq.spham}
\eeq
which contains an anisotropy constant $\tilde{D}$ and a magnetic field $\tilde{\bm{H}}=(\tilde{H}_x,0,\tilde{H}_z)$. 
Through this paper, we set $g\mu_B=1$. 
Owing to the uniaxial anisotropy, the system exhibits magnetic hysteresis. 
For systems with finite $S$, i.e. with the discrete energy levels, magnetic hysteresis is also discrete and is associated with dynamical jumps. 

The classical correspondence of the Hamiltonian (\ref{Eq.spham}) is nothing but the energy expression of the Stoner-Wohlfarth model~\cite{SW} 
\begin{equation}
E=-Dm_z^2-H_xm_x-H_zm_z,
\label{SWene}
\end{equation}
where $\bm{m}=(m_x,m_y,m_z)$ is a classical spin described by a unit vector, $|\bm{m}|=1$. 
In the past, this model was devoted to the study of the magnetization reversal of single-domain ferromagnetic nano-particles under a tilted magnetic field. 
The important relation between the quantities $\tilde{D}$ and $D$ will be given in the next subsection. 

\begin{figure}
\begin{tabular}{rl}
(a) & \vspace{-1cm} \\
(b) & \vspace{-1cm}\includegraphics[width=7.5cm]{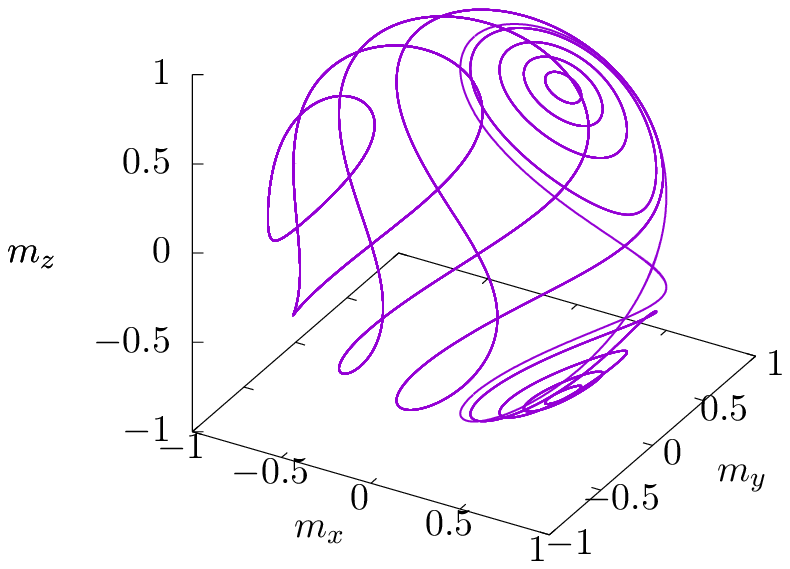} \\
(c) & \vspace{-1cm}\includegraphics[width=7.5cm]{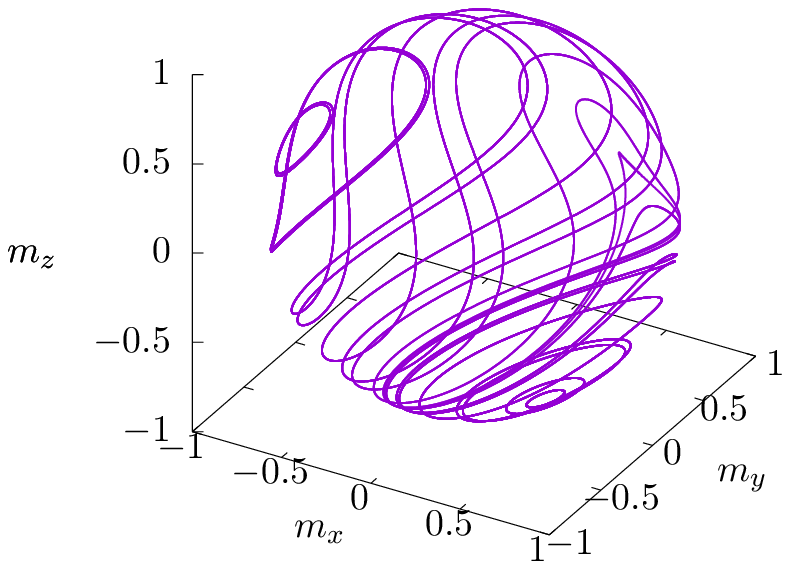} \\
 & \includegraphics[width=7.5cm]{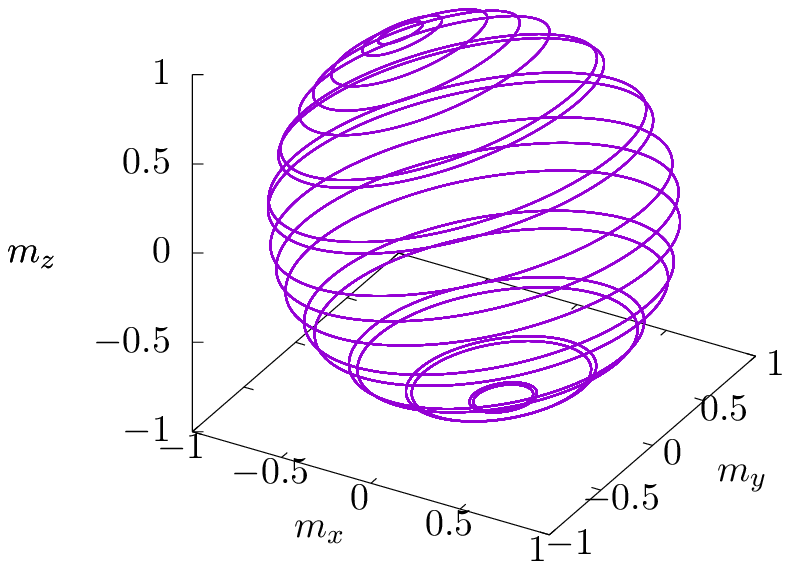}
\end{tabular}
\caption{\label{SWtrajectory}Typical trajectories for the cases (a) with the degenerated ground states, $(2D)^{2/3}>(H_x)^{2/3}+(H_z)^{2/3}$, $H_z=0$, (b) at the Stoner-Wohlfarth point, $(2D)^{2/3}=(H_x)^{2/3}+(H_z)^{2/3}$, $H_z=H_\mathrm{SW}$, and
(c) without the metastable state, $(2D)^{2/3}<(H_x)^{2/3}+(H_z)^{2/3}$, $H_z=-4$. Here, $D=1$ and $H_x=1$. }
\end{figure}

If the magnetic field is applied along a direction in the hemisphere opposite to spontaneous magnetization, the latter classical model (\ref{SWene}) can be in metastable equilibrium. 
When the magnetic field becomes large so that the energy barrier vanishes, the system changes from metastable to unstable. 
Such a point associated with a metastable-to-stable transition is generally called the spinodal point. 
In the present model, it is called the Stoner-Wohlfarth point and given by
\begin{equation}
(2D)^{2/3}=(H_x)^{2/3}+(H_z)^{2/3}.
\end{equation}
This equation represents the famous Stoner-Wolhfart astroid which was actually observed in experiment~\cite{doi:10.1063/1.364656}. 

In Fig.~\ref{SWtrajectory}, we show the three main types of trajectories (a) with the metastable state, inside the Stoner-Wohlfarth astroid $(2D)^{2/3}>(H_x)^{2/3}+(H_z)^{2/3}$, (b) at the Stoner-Wohlfarth point, on the astroid $(2D)^{2/3}=(H_x)^{2/3}+(H_z)^{2/3}$, and (c) without the metastable state, outside the astroid $(2D)^{2/3}<(H_x)^{2/3}+(H_z)^{2/3}$.

\subsection{Quantum Stoner-Wohlfarth model}
When we study the quantum effects of the Stoner-Wohlfarth model, it is convenient to adopt the quantum Stoner-Wohlfarth model rather than the spin Hamiltonian (\ref{Eq.spham}). 
In the quantum Stoner-Wohlfarth model, we replaced the classical spin in Eq.~(\ref{SWene}) by the normalized quantum spin operator
\beq
\bm{s}=\bm{S}/S,
\eeq 
which satisfies the following commutation relation
\begin{equation}
[s_i,s_j]=\frac{i}{S}\epsilon_{ijk}s_k.
\end{equation}
Here and hereafter, we take $\hbar=1$ and obtain the quantum Stoner-Wohlfath Hamiltonian
\begin{equation}
\mathcal{H}(t)=-Ds_z^2-H_xs_x-H_z(t)s_z,
\label{QSWham}
\end{equation}
where the explicit time-dependence of the longitudinal field $H_z(t)$, which is taken in our analyses, is given by
\begin{equation}
H_z(t)=H_z(0)-ct,
\end{equation}
where $c$ is a given sweeping rate. 

In order to make clear the relation between the spin Hamiltonian (\ref{Eq.spham}) and the quantum Stoner-Wohlfarth model (\ref{QSWham}), we rewrite Eq.~(\ref{Eq.spham}) as
\beq
\tilde{\mathcal{H}}=S(-\tilde{D}Ss_z^2-\tilde{H}_xs_x-\tilde{H}_zs_z).
\eeq
Comparing with Eq. (\ref{QSWham}), we find the ralations $D=\tilde{D}S$, $\bm{H}=\tilde{\bm{H}}$, and $\mathcal{H}=\tilde{\mathcal{H}}/S$. 
The overall factor $S$ causes the following time rescaling
\beq
\tau=\frac{t}{S}, 
\eeq
which gives
\begin{equation}
H_z(t)=H_z(0)-ct=H_z(0)-v\tau,\quad v=cS,
\label{Eq.sweeping}
\end{equation}
where $v$ is the sweeping rate in the spin Hamiltonian (\ref{Eq.spham}). 
In the rest of the present paper, we shall consider the Hamiltonian (\ref{QSWham}), which is the same as in our previous paper~\cite{HBM}. 
Note that the corresponding anisotropy constant is $D=\tilde{D}S$, where $\tilde{D}$ is the usual one in the spin Hamiltonian (\ref{Eq.spham}). 

\subsection{Population dynamics}

\begin{figure}
\begin{tabular}{rl}
(a) & \\
(b) & \includegraphics[width=8cm]{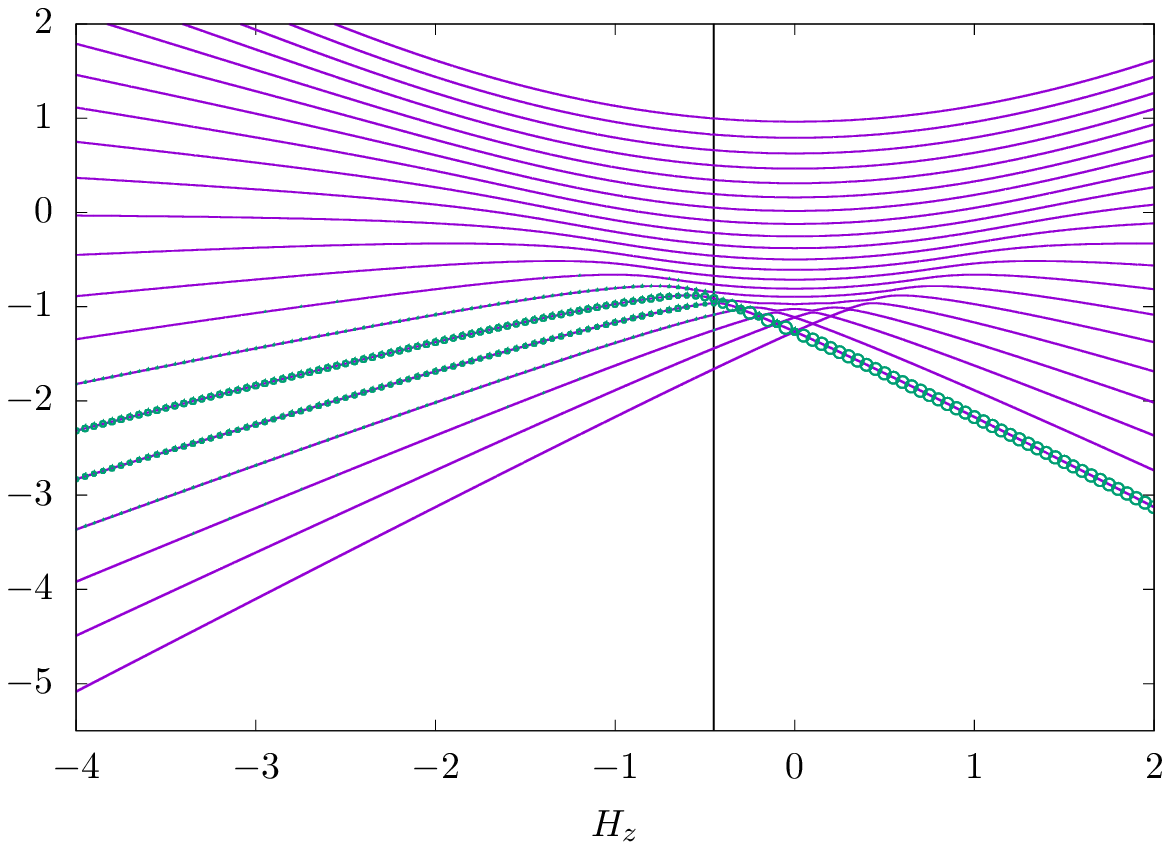} \\
 & \includegraphics[width=8cm]{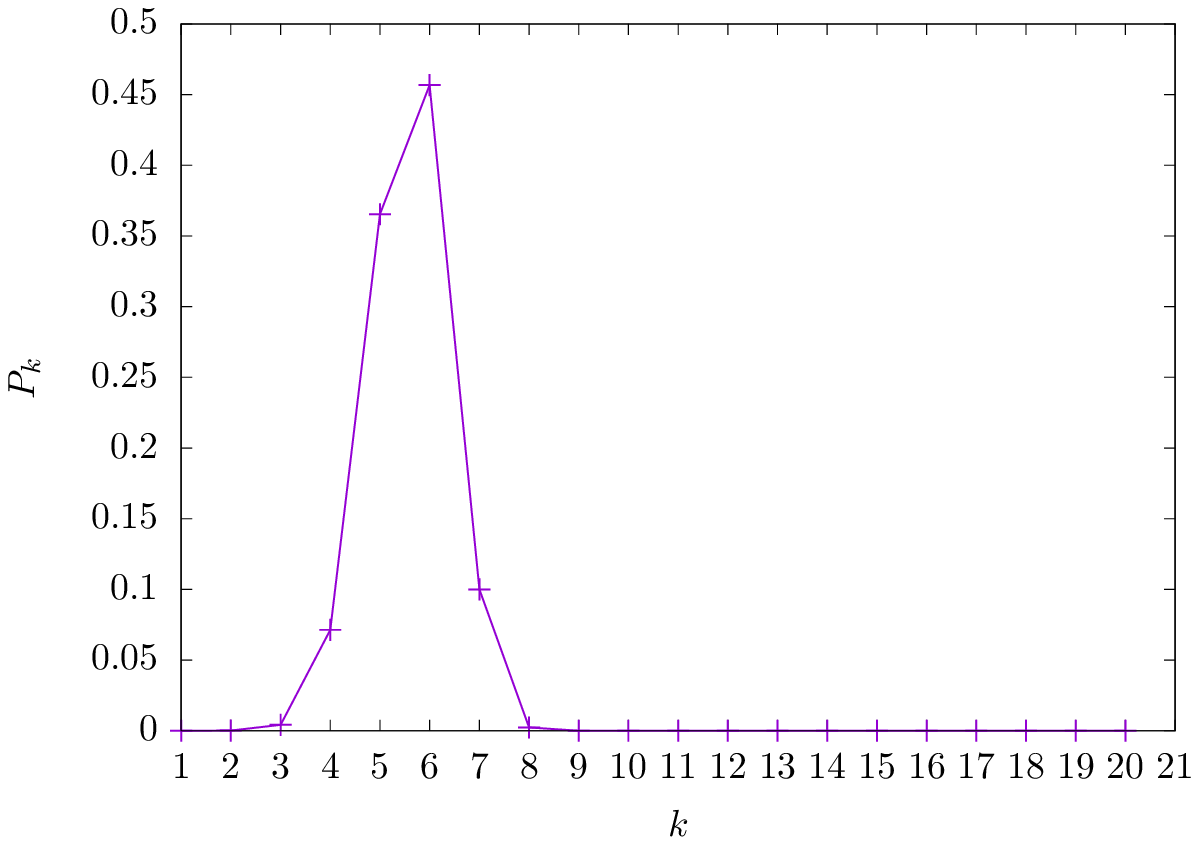}
\end{tabular}
\caption{\label{s10fig}(a) Energy spectrum of the quantum Stoner-Wohlfarth model for spin-size $S=10$, the anisotropic constant $D=1$, and the transverse field $H_x=1$. 
The horizontal axis is the longitudinal field $H_z$. The purple curves represent the eigenenergies. Population dynamics under the swept field with the velocity $v=0.05$ is represented by the green circles. The black line denotes the Stoner-Wohlfarth point.
(b) The distribution of the populations at $H_z(t)=-4$.  The horizontal axis represents the number of the state $k$ 
and the vertical axis represents the populations $P_k$. }
\end{figure}

Let $\{|\psi_k(t)\rangle\}$ be the instantaneous eigenstates of the Hamiltonian (\ref{QSWham}), 
\beq
\mathcal{H}(t)|\psi_k(t)\rangle=E_k(H_z(t))|\psi_k(t)\rangle.
\eeq
Under the swept field (\ref{Eq.sweeping}), the time-evolution is given by the time-dependent Schr\"odinger equation
\begin{eqnarray}
i\frac{\partial}{\partial t}|\psi(t)\rangle&=&\mathcal{H}(t)|\psi(t)\rangle, \\
|\psi(t)\rangle&=&\sum_kc_k(t)|\psi_k(t)\rangle,
\end{eqnarray}
where $c_k(t)$ is the time-dependent coefficient. 
Now, we introduce the population of the eigenstate
\beq
P_k(t)=|c_k(t)|^2.
\eeq
A typical energy spectrum given by the ensemble $\{E_k(H_z(t))\}_{k=1, \cdots, 2S+1}$ is depicted in Fig.~\ref{s10fig} (a) for $S=10$, where the $(2S+1=21)$ eigenenergies are plotted as a function of $H_z(t)$. 
In this figure, the population dynamics is simulated for the initial condition $H_z(0)=2$ with the sweeping rate $v=0.05$, starting from the ground state. 
The size of circles denotes the population of the eigenstates $P_k(t)$. 
The distribution of the eigenstate populations at $H_z(t)=-4$ is depicted in Fig.~\ref{s10fig} (b). 

\begin{figure}
\includegraphics[width=8cm]{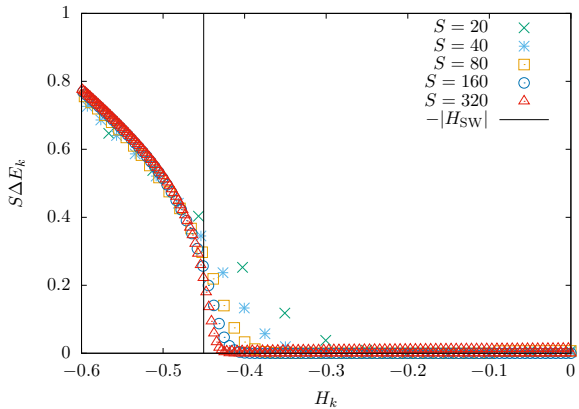}
\caption{\label{gapfig}Critical behavior of gaps at avoided crossings on the metastable branch, which and its finite-size scaling are found by Hatomura \textit{et al.} [Phys. Rev. Lett. \textbf{116}, 037203 (2016)]. Gaps are plotted from spin-size $S=20$ to $S=320$. The anisotropic constant is $D=1$ and the transverse field is $H_x=1$. The vertical axis is the scaled gaps $S\Delta E_k$ and the horizontal axis is the corresponding longitudinal magnetic fields $H_k$. The Stoner-Wohlfarth point $H_\mathrm{SW}$ is represented by the black line. }
\end{figure}

The scattering process is understood as follows. 
When the longitudinal field $H_z(t)$ is swept from positive to negative, the fully occupied initial state is scattered at the successive anti-crossing points with the states corresponding to the magnetization $M_z=-S,-S+1,\cdots ,+S-1$ along the diadiabatic continuations of the initial state (metastable branch). 
The population $P_k(t)$ at a certain negative magnetic field $H_z(t)$ can be interpreted as the amount of the scattering at the $k$th avoided-crossing point. 
Here, the $k$th avoided-crossing point is the anti-crossing point of the states corresponding to $M_z=+S$ and the state corresponding to $M_z=-S+k-1$. 
The energy gap of the $k$th avoided-crossing point is denoted by $\Delta E_k$, and the corresponding field is defined as $H_k$, which gives the minimum energy gap at the anti-crossing level. 
The amount of the scattering is, of course, related to the size of the energy gap. 
In the classical limit $S\to\infty$, the scaled gap $S\Delta E_k$ depicted in Fig.~\ref{gapfig} is responsible to the scattering phenomenon. 
For $|H_z|<|H_\mathrm{SW}|$, the state remains in the metastable state, which indicates $S\Delta E_k$ vanishes while it has a finite value for $|H_z|>|H_\mathrm{SW}|$. 
This situation is responsible for the critical behavior at the Stoner-Wohlfarth point, which is characterized by the scaling law of the renormalized gaps $S\Delta E_k$ versus the longitudinal field $H_k$ observed in the previous paper~\cite{HBM}
\beq
(S\Delta E_k)^2=S^{-1/3}g((H_k-H_\mathrm{SW})S^{2/3}),
\label{finitesizeform}
\eeq
where $g(\cdot)$ is the scaling function. 
This scaling function turned out to be identical to one of the spinodal critical scaling~\cite{PhysRevE.81.011135}. 

In the present paper, we study the distribution of the populations $P_k(t)$ at the field 
$H_z(t)=-4$, $P_k\equiv P_k(t)|_{H_k(t)=-4}$, after the scattering region, i.e. far beyond the Stoner-Wohlfarth point (Fig.~\ref{s10fig} (a)). 

%
%
\section{\label{distribution}Distribution of populations}
\subsection{Distribution and its scaling behavior}
\begin{figure}
\begin{tabular}{rl}
(a) & \\
(b) & \includegraphics[width=8cm]{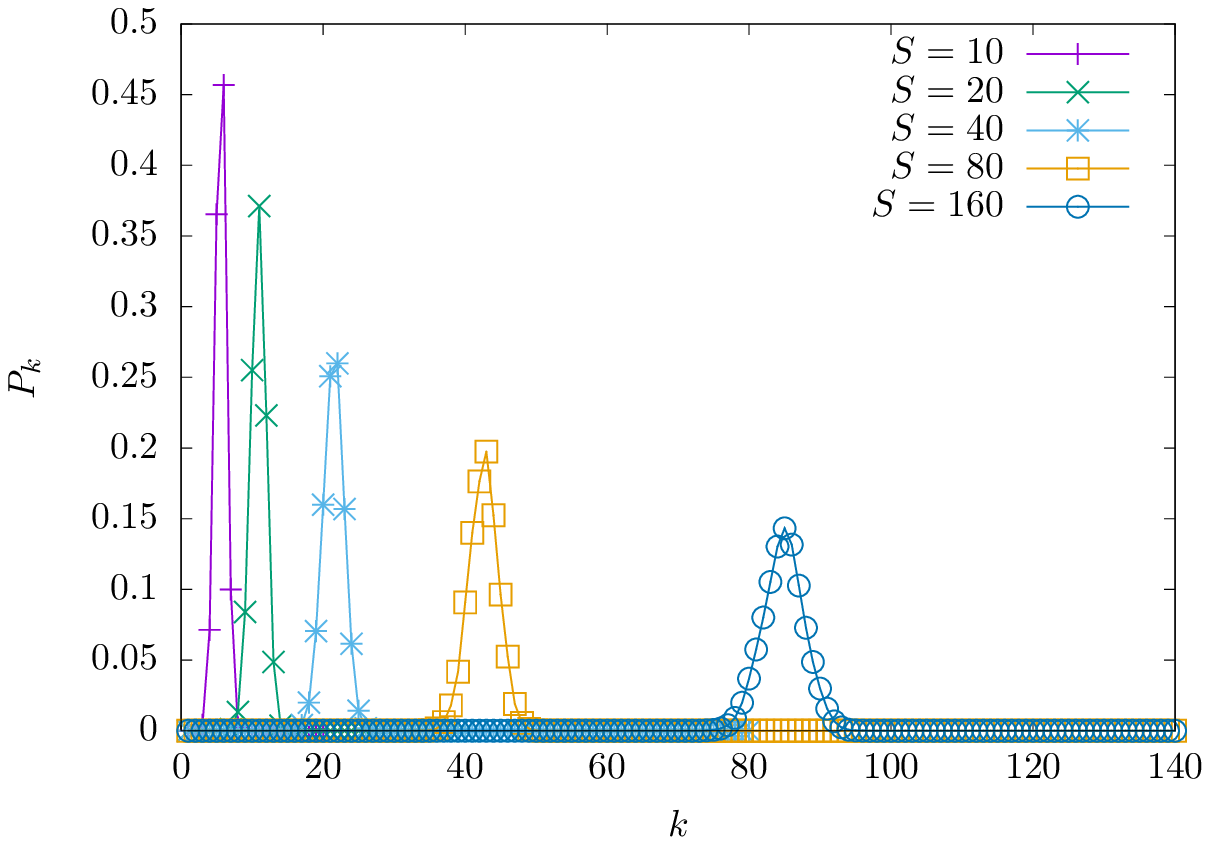} \\
 & \includegraphics[width=8cm]{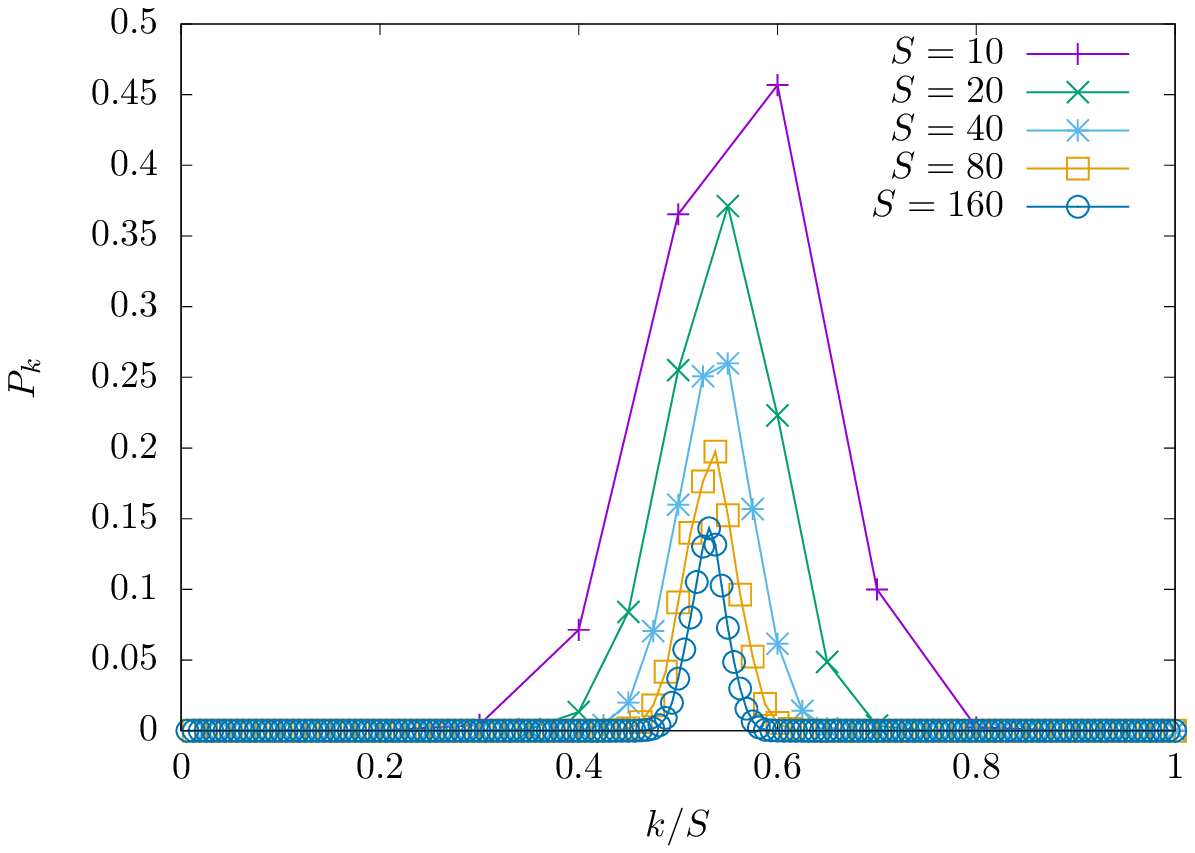} 
\end{tabular}
\caption{\label{Fig.Pkk}
(a)
The distributions of populations for
$S=10,20,40,80,$ and 160. 
(b)
The data are plotted as a function of the scaled parameter $k/S$.
}
\end{figure}

\begin{figure}
$$
\includegraphics[width=8cm]{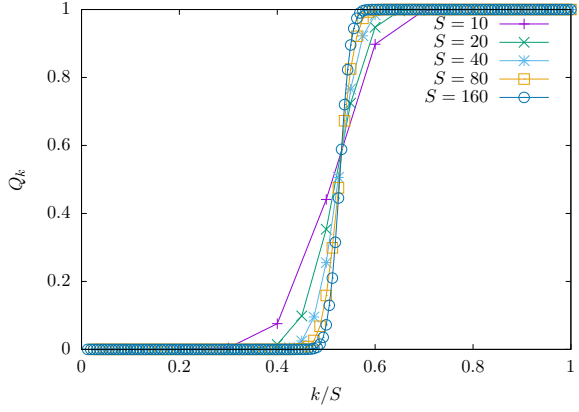}
$$
\caption{\label{Fig.APkk}
The accumulated populations for
$S=10,20,40,80,$ and 160. 
The horizontal axis is the scaled parameter $k/S$ and the vertical axis is the accumulated populations $Q_k$. 
}
\end{figure}

First, we study $S$ dependence of the eigenstate populations $\{P_k\}$. 
The distributions are plotted versus $k$ in Fig.~\ref{Fig.Pkk} (a) for spin $S=10$ to $160$. 
They are given again in Fig.~\ref{Fig.Pkk} (b) versus $k/S$. 
This leads to a concentration of the curves at a certain position. 
In Fig.~\ref{Fig.APkk}, we show the accumulated population 
\beq
Q_k\equiv \sum_{k'=1}^kP_{k'}
\label{APkk}
\eeq
as a function of $k/S$. 
Here, we find that the accumulated populations reach almost 1 around $k/S\approx0.7$, and thus the scattering processes almost finish around there. 
Indeed, the populations are almost zero for large $k$, which do not contribute to the scattering processes. 
Therefore, hereafter, we will not care about large $k$ regions. 
In addition, we find that all the curves with different $S$ cross at almost the same point. 
We denote this crossing point as $k_\mathrm{peak}^\mathrm{cl}/S$. 

\begin{figure}
\includegraphics[width=8cm]{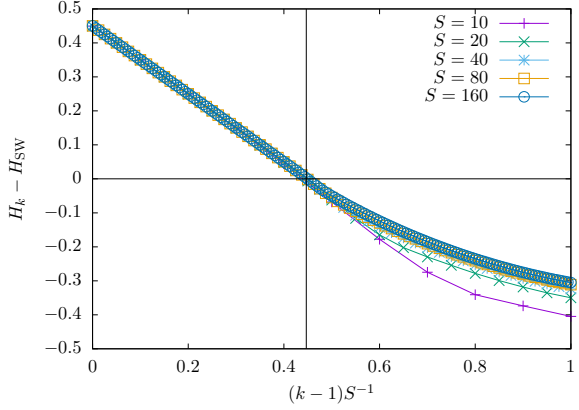}
\caption{\label{kdephk}Relations between $H_k$ and $k$. From the first avoided crossing to the Stoner-Wohlfarth point, $H_\mathrm{SW}<H_k<0$, $S$ dependence of the field corresponding to the $k$th gap is given by $H_k\propto kS^{-1}$. The horizontal black line represents the Stoner-Wohlfarth point and the vertical black line is depicted to estimate corresponding $k$. }
\end{figure}

When we consider the distributions, the field $H_k$ is more meaningful than the label $k$, and so we plot the difference of the fields $H_k-H_\mathrm{SW}$ as a function of $(k-1)/S$ in Fig.~\ref{kdephk}. 
Here, we estimate the Stoner-Wohlfarth point in terms of $k$ for later analysis, and it is given by $k_\mathrm{SW}\simeq1+0.45S$. 
We find this is actually a good scaling until the Stoner-Wohlfarth point. 
However, this scaling is broken beyond the Stoner-Wohlfarth point. 
This point will be discussed in more details later. 

\begin{figure}
\includegraphics[width=8cm]{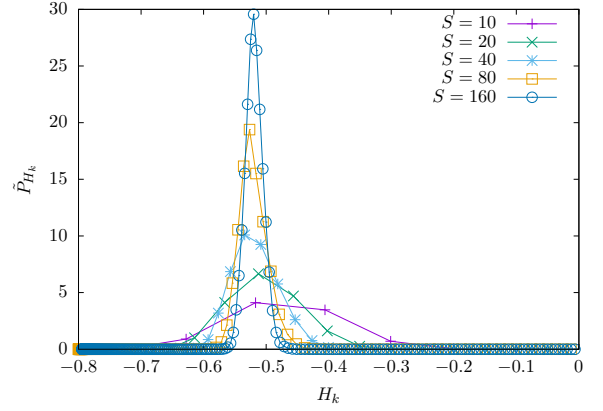}
\caption{\label{PHzHz}
The distributions $\tilde{P}_{H_k}$ as a function of $H_k$ for
$S=10,20,40,80$ and 160. 
The horizontal axis is the fields $H_k$ and the vertical axis is the populations $\tilde{P}_{H_k}$. 
}
\end{figure}

\begin{figure}
\includegraphics[width=8cm]{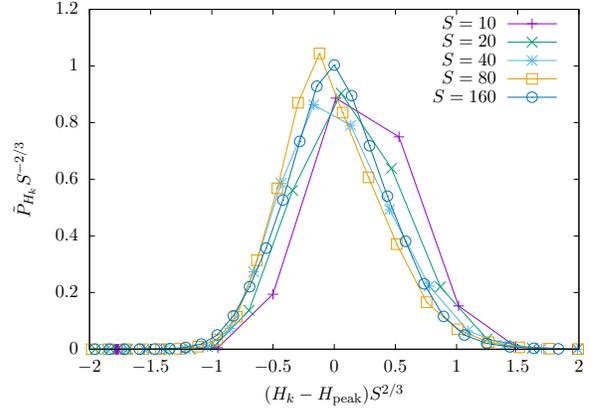}
\caption{\label{SPHzHz}
The possible scaling plot for $\tilde{P}_{H_k}$. 
The horizontal axis is scaled as $(H_k-H_\mathrm{peak})S^{2/3}$ and the vertical axis is scaled as $\tilde{P}_{H_k}S^{-2/3}$. 
}
\end{figure}

Now, we study the population $\tilde{P}_{H_k}$, which is a function of $H_k$, measuring the scattering rate at the field $H_k$ during the sweeping process. 
The relation between the populations $P_k$ and $\tilde{P}_{H_k}$ is given by
\beq
P_k\Delta k=\tilde{P}_{H_k}\Delta H_k\quad\to\quad\tilde{P}_{H_k}=P_k\left(\frac{\Delta H_k}{\Delta k}\right)^{-1},
\label{Eq.trans}
\eeq
where $\Delta k=k-(k-1)=1$ and $\Delta H_k=H_k-H_{k-1}$. 
By using this relation, we obtain the distributions of the populations $\tilde{P}_{H_k}$ as depicted in Fig.~\ref{PHzHz}. 
In this expression, the areas of the distributions are conserved and equal to unity. 
Furthermore, the peak positions are almost $S$ independent for large $S$  and nearly given by
\beq
H_{\rm peak}\simeq 0.52, 
\eeq
suggesting the existence of a scaling plot. 
In order to keep the conservation law for the areas of the distributions, the scaling plot must take the form, $(\tilde{P}_{H_k}S^{-\alpha},(H_k-H_\mathrm{peak})S^\alpha)$, where $\alpha$ is a real number. 
Taking $\alpha=2/3$, we obtain the scaled distributions as shown in Fig.~\ref{SPHzHz}. 
This is actually the scaling law of the distributions of the populations for large $S$. 

\begin{figure}
\includegraphics[width=8cm]{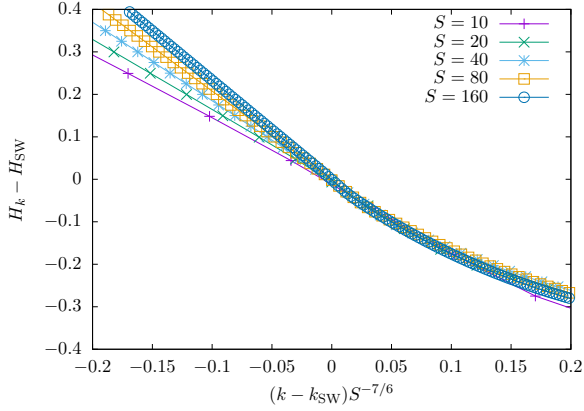}
\caption{\label{Fig.int2}Relation between $H_k$ and $k$ above the Stoner-Wohlfarth point, $H_k<H_\mathrm{SW}<0$. The horizontal axis is $(k-k_\mathrm{SW})S^{-7/6}$ and the vertical axis is $H_k-H_\mathrm{SW}$. }
\end{figure}

\begin{figure}
\includegraphics[width=8cm]{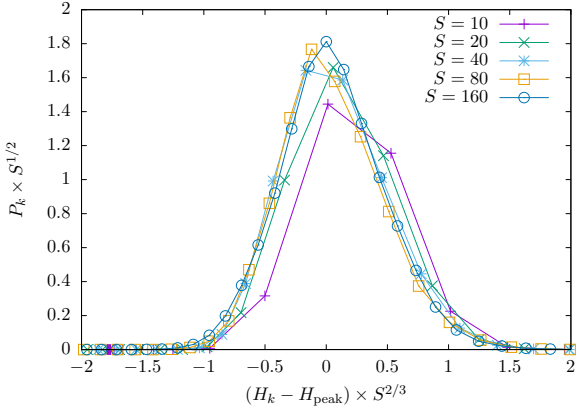}
\caption{\label{scaledistfig}The possible scaling for distributions. The horizontal axis is plotted as $(H_k-H_\mathrm{peak})S^{2/3}$ and the vertical axis is $P_kS^{1/2}$. }
\end{figure}

In the above calculations, we adopted finite differences for the transformation from $k$ to $H_k$. 
This causes non-negligible errors, especially for small $S$. 
Indeed, the results are quantitatively different when we adopt $\Delta H_k=H_{k+1}-H_k$ instead of $H_k-H_{k-1}$. 
Therefore, we consider the continuous limit $S\to\infty$ in Eq.~\ref{Eq.trans}, 
\beq
\tilde{P}_{H_k}=P_k\left(\frac{\Delta H_k}{\Delta k}\right)^{-1}\to\quad\tilde{P}(H_k)=P(k)\left(\frac{dH(k)}{dk}\right)^{-1},
\eeq
where $\tilde{P}(H_k)$ and $P(k)$ are the continuous limit of $\tilde{P}_{H_k}$ and $P_k$, and $H(k)$ is the continuous limit of $H_k$. 
From Fig.~\ref{kdephk}, the derivative $dH(k)/dk$ is proportional to $S^{-1}$ for $0<|H(k)|<|H_\mathrm{SW}|$. 
Now, we consider the scaling for $|H(k)|>|H_\mathrm{SW}|$. 
The possible scaling is plotted in Fig.~\ref{Fig.int2}. 
From this scaling, the derivative $dH(k)/dk$ is proportional to $S^{-7/6}$ for $|H(k)|>|H_\mathrm{SW}|$. 
Therefore, the scaling property of the vertical axis is $\tilde{P}(H_k)S^{-2/3}\propto P(k)S^{1/2}$ above the Stoner-Wohlfarth point, and thus we obtain the scaled distributions as shown in Fig.~\ref{scaledistfig}. 
We remark that the heights of the distributions are different from Fig.~\ref{SPHzHz} because we neglect the coefficient of $dH(k)/dk\propto S^{-7/6}$. 

\subsection{Velocity dependence of the peak fields}
\begin{figure}
\includegraphics[width=8cm]{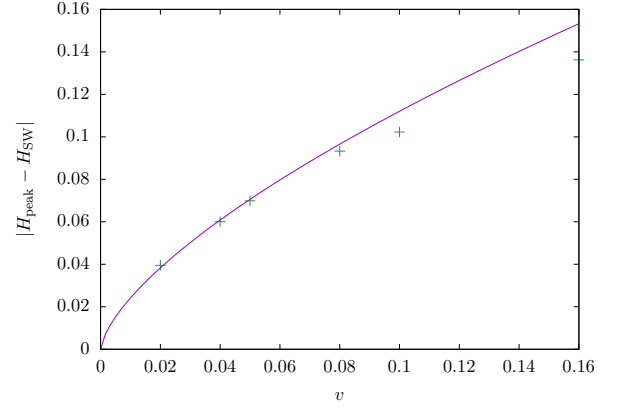}
\caption{\label{vdeppeakfig}Sweeping-rate dependence of the peaks of the distributions. Spin-size is $S=160$, the anisotropic constant is $D=1$, and the transverse field is $H_x=1$. The horizontal axis represents the sweeping rate $v$ and the vertical axis represents the difference of the peak field and the Stoner-Wohlfarth point $|H_\mathrm{peak}-H_\mathrm{SW}|$. The purple curve is $|H_\mathrm{peak}-H_\mathrm{SW}|=0.52v^{2/3}$. 
}
\end{figure}

In the above subsection, we studied the $S$ dependence of the distribution for the particular sweeping rate $v=0.05$. 
In this subsection, we investigate the sweeping rate dependence of the shift $\Delta H_\mathrm{peak}=|H_\mathrm{peak}-H_\mathrm{SW}|$ for the particular spin $S=160$. 
The result of the calculations is given in Fig.~\ref{vdeppeakfig}. 
Here, we find that the shift $\Delta H_\mathrm{peak}$ is proportional to $v^{2/3}$ for small velocities $v\lesssim0.08$. 
This dependence can be understood from the scaled energy gaps depicted in Fig.~\ref{gapfig}  and the scaling property (\ref{finitesizeform}). 
Indeed, from the scaling form (\ref{finitesizeform}) and Fig.~\ref{gapfig}, we find asymptotically
\beq
(S\Delta E_k)^2\propto |H_k-H_{\rm SW}|^{1/2},
\eeq
beyond the Stoner-Wohlfarth point $|H_k|>|H_\mathrm{SW}|$ for spin $S\to\infty$. 
The amount of the scattered populations in a given small interval is approximately proportional to the adiabaticity parameter $\delta\equiv\sum_k(\Delta E_k)^2/c$ when the avoided-crossing gaps $\Delta E_k$ are small and well-separated. 
Here, summation is taken over a given interval of $H_z$ and $c$ is the sweeping rate. 
Thus, the amount of the scattered populations in the interval $[H_\mathrm{peak},H_\mathrm{SW}]$ is proportional to
\begin{eqnarray}
\sum_{k=k_\mathrm{SW}}^{k_\mathrm{peak}^\mathrm{cl}}\frac{(\Delta E_k)^2}{c}&=&\sum_{k=k_\mathrm{SW}}^{k_\mathrm{peak}^\mathrm{cl}}\frac{1}{S}\frac{(S\Delta E_k)^2}{v} \notag \\
&\propto&\sum_{k=k_\mathrm{SW}}^{k_\mathrm{peak}^\mathrm{cl}}\frac{1}{S}\frac{|H_k-H_\mathrm{SW}|^{1/2}}{v},
\end{eqnarray}
for large $S$. 
Although $S$ dependence of $dH(k)/dk$ is complicated and $k_\mathrm{SW}/S$ is also $S$ dependent, we regard the summation as the following integral
\begin{eqnarray}
\sum_{k=k_\mathrm{SW}}^{k_\mathrm{peak}^\mathrm{cl}}\frac{1}{S}\frac{|H_k-H_\mathrm{SW}|^{1/2}}{v}&\simeq&\int_{H_\mathrm{SW}}^{H_\mathrm{peak}}\frac{|H-H_\mathrm{SW}|^{1/2}}{v}dH \notag \\
&\propto&\frac{|H_\mathrm{peak}-H_\mathrm{SW}|^{3/2}}{v},
\end{eqnarray}
for large $S$ and just above the Stoner-Wohlfarth point. 
As the scattered populations should be invariant for various velocities $v$, the $v$ dependence of the peak $H_\mathrm{peak}$ is given by
\beq
|H_\mathrm{peak}-H_\mathrm{SW}|\propto v^{2/3}. 
\label{peakvdep}
\eeq
It is obvious that this behavior (\ref{peakvdep}) does not work for large velocities $v$ because the above assumptions will not hold for large fields $|H-H_\mathrm{SW}|$. 

%
%
\section{\label{dissipative}Effects of the environment \\ on the beating dynamics}
\subsection{Beating phenomenon}
\begin{figure*}
\begin{tabular}{cc}
\includegraphics[width=8cm]{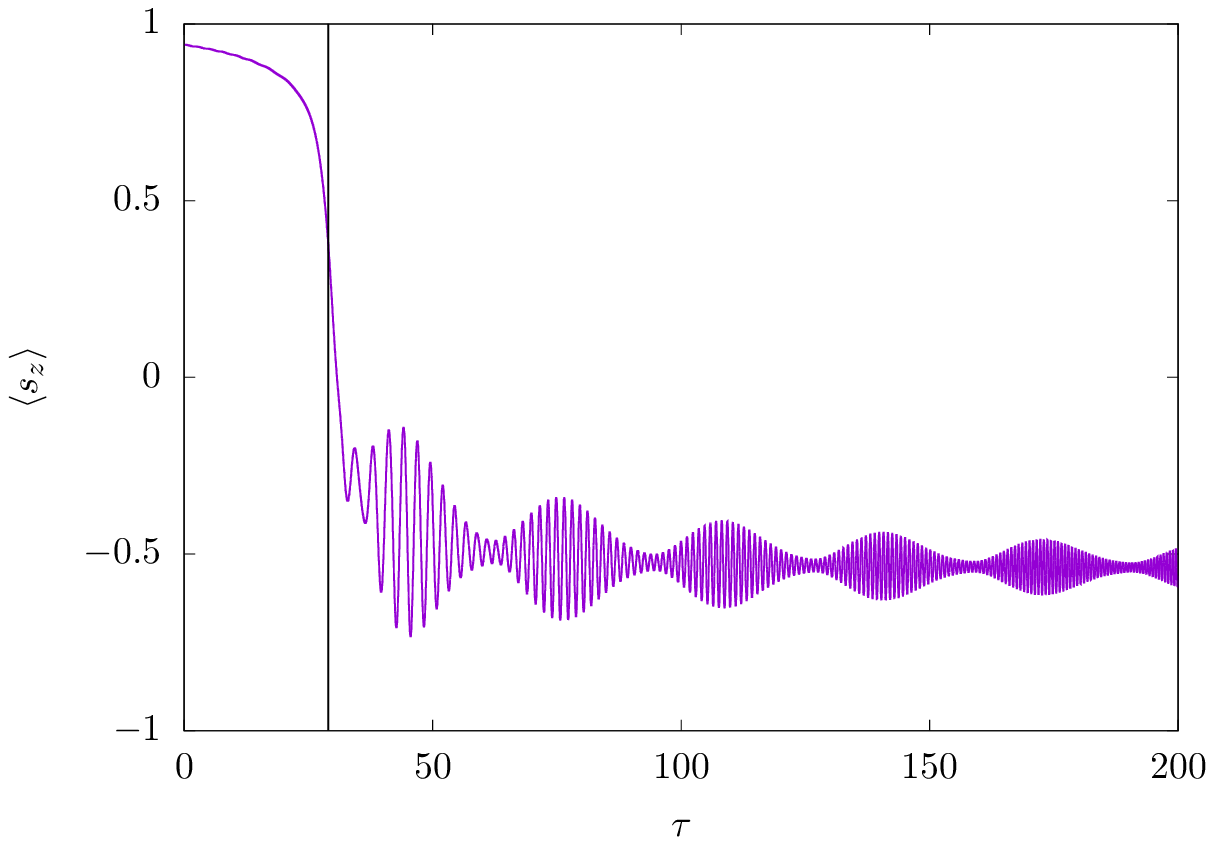} & \includegraphics[width=8cm]{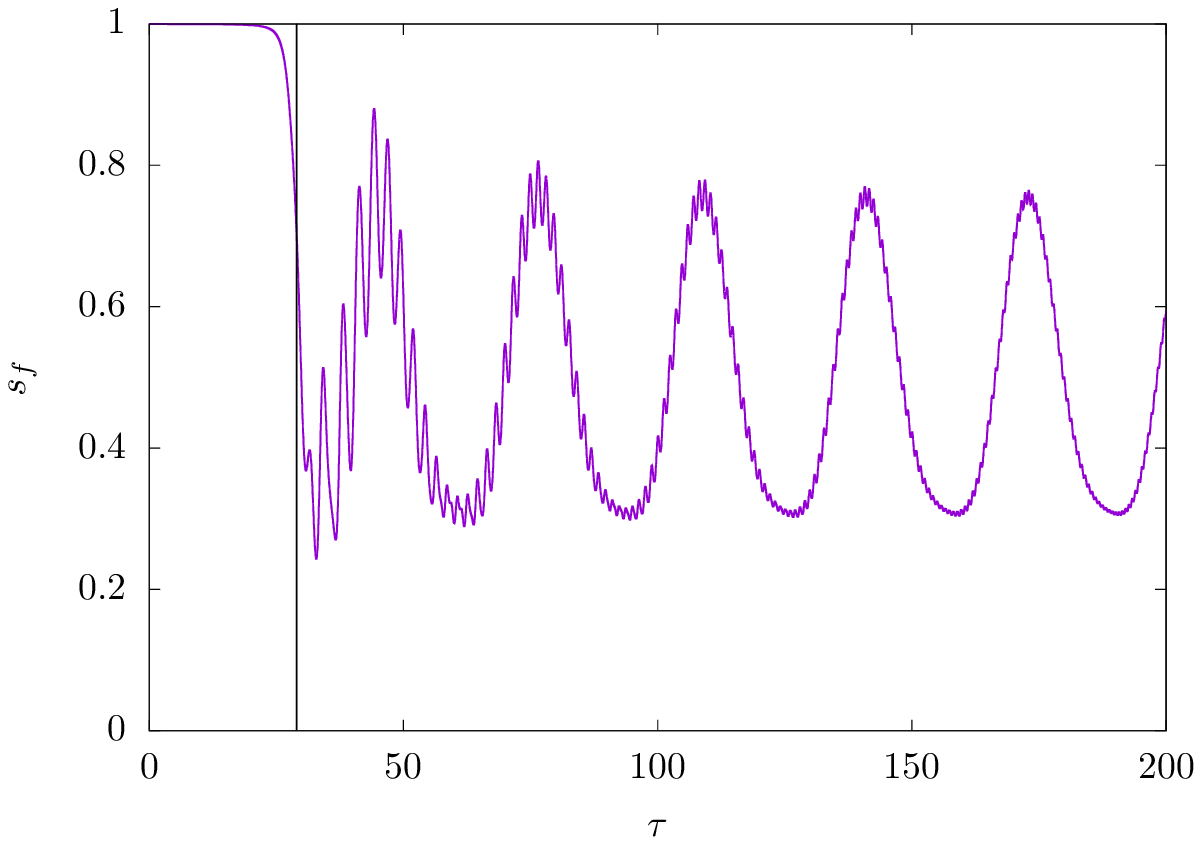} \\
\includegraphics[width=8cm]{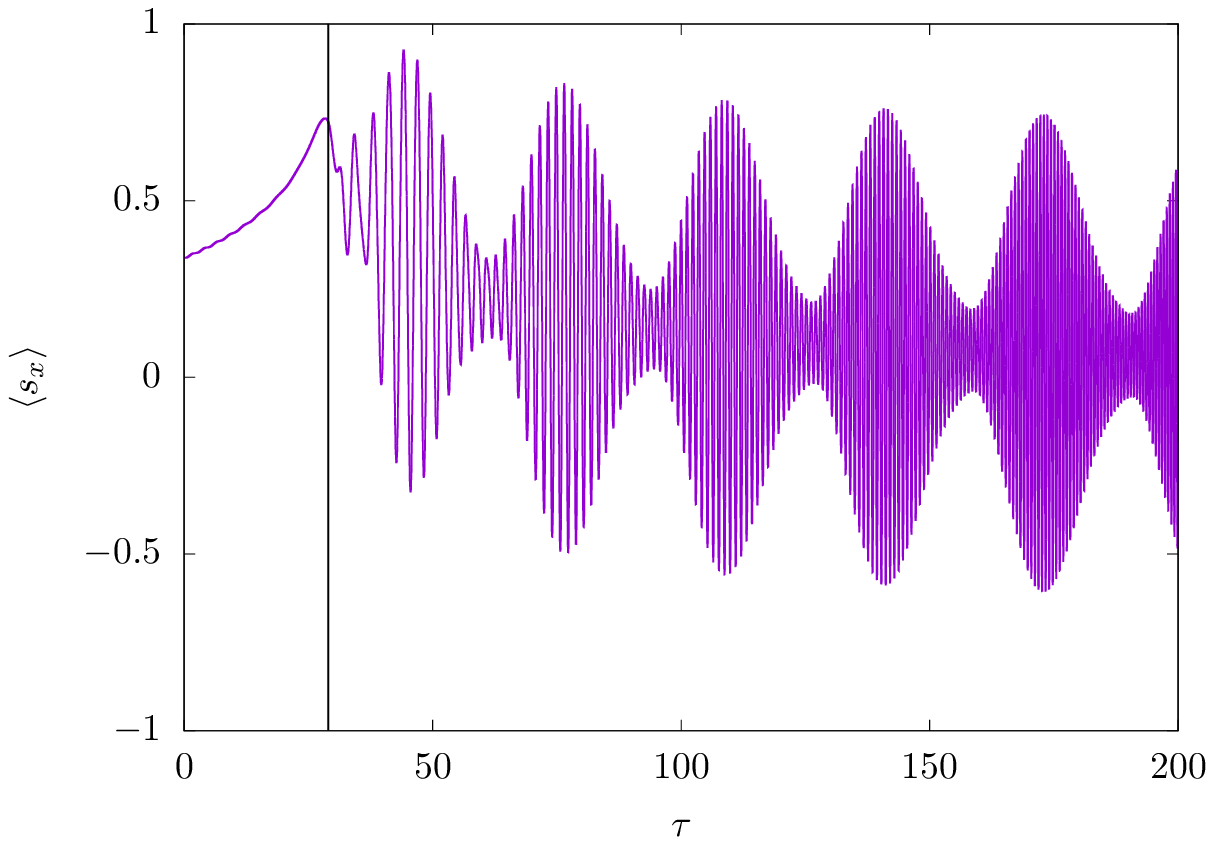} & \includegraphics[width=8cm]{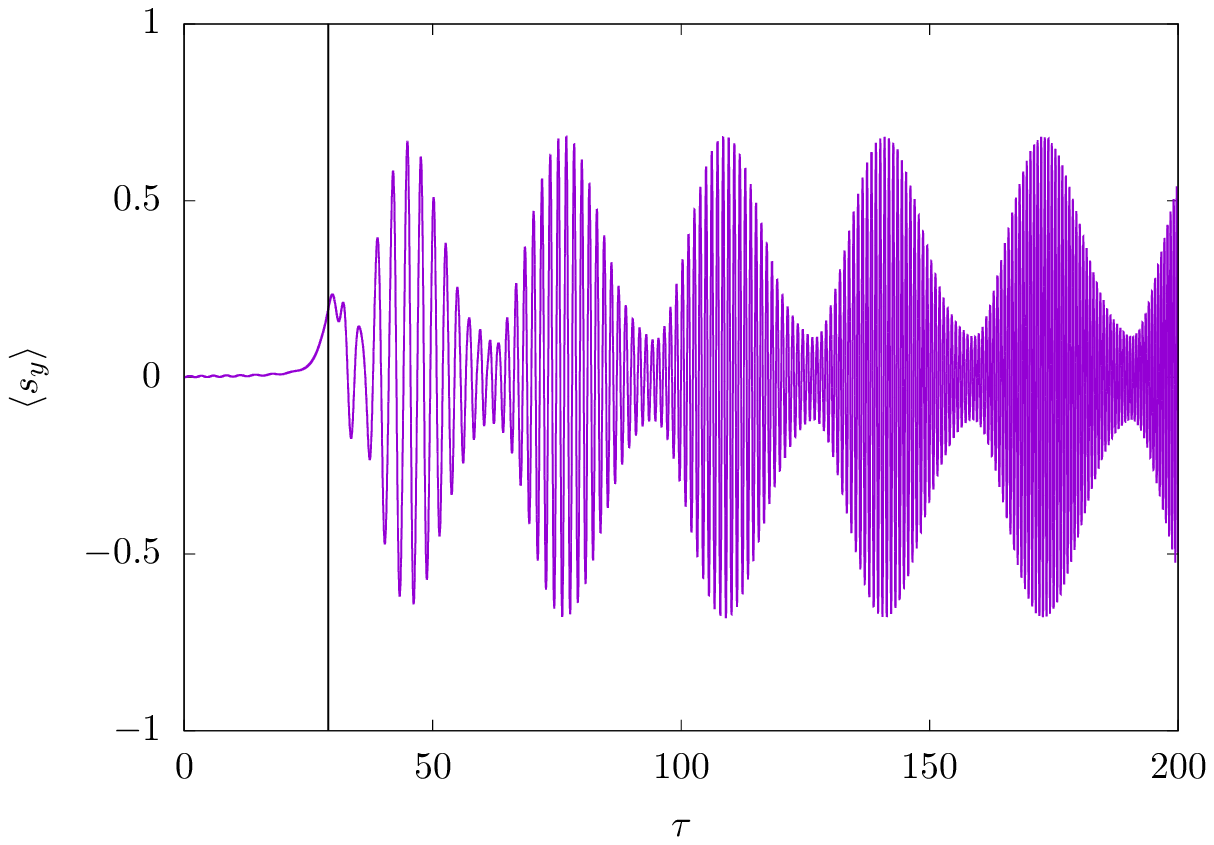} 
\end{tabular}
\caption{\label{beatfig}Beatings of the spin-components and the spin-fidelity. The horizontal axis is time $\tau$ and the vertical axes are (top left) $\langle s_z\rangle$, (top right) $s_f$, (bottom left) $\langle s_x\rangle$, and (bottom right) $\langle s_y\rangle$. Spin-size is $S=10$, the anisotropic constant is $D=1$, and the transverse field is $H_x=1$. The longitudinal field is swept from $H_z(0)=1$ with the sweeping rate $v=0.05$. }
\end{figure*}
\begin{figure*}
\begin{tabular}{cc}
\includegraphics[width=8cm]{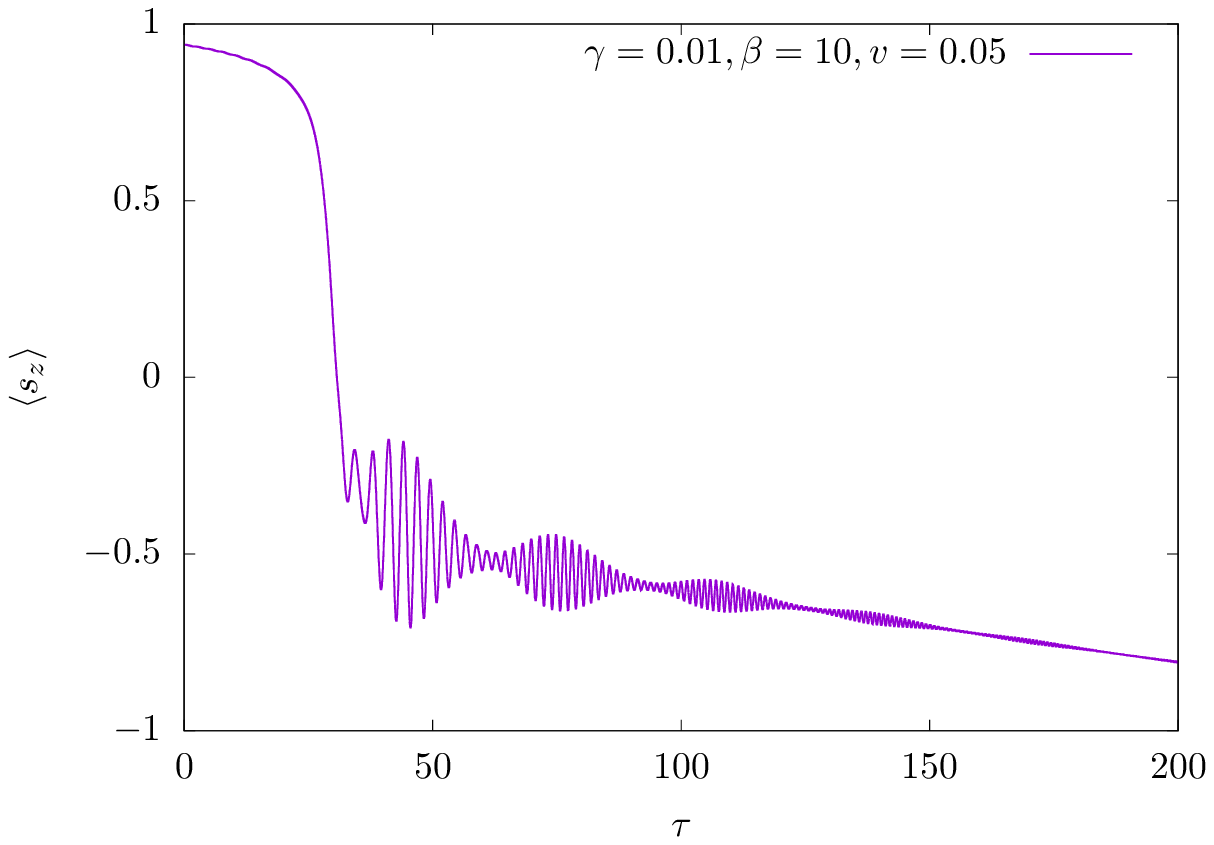} & \includegraphics[width=8cm]{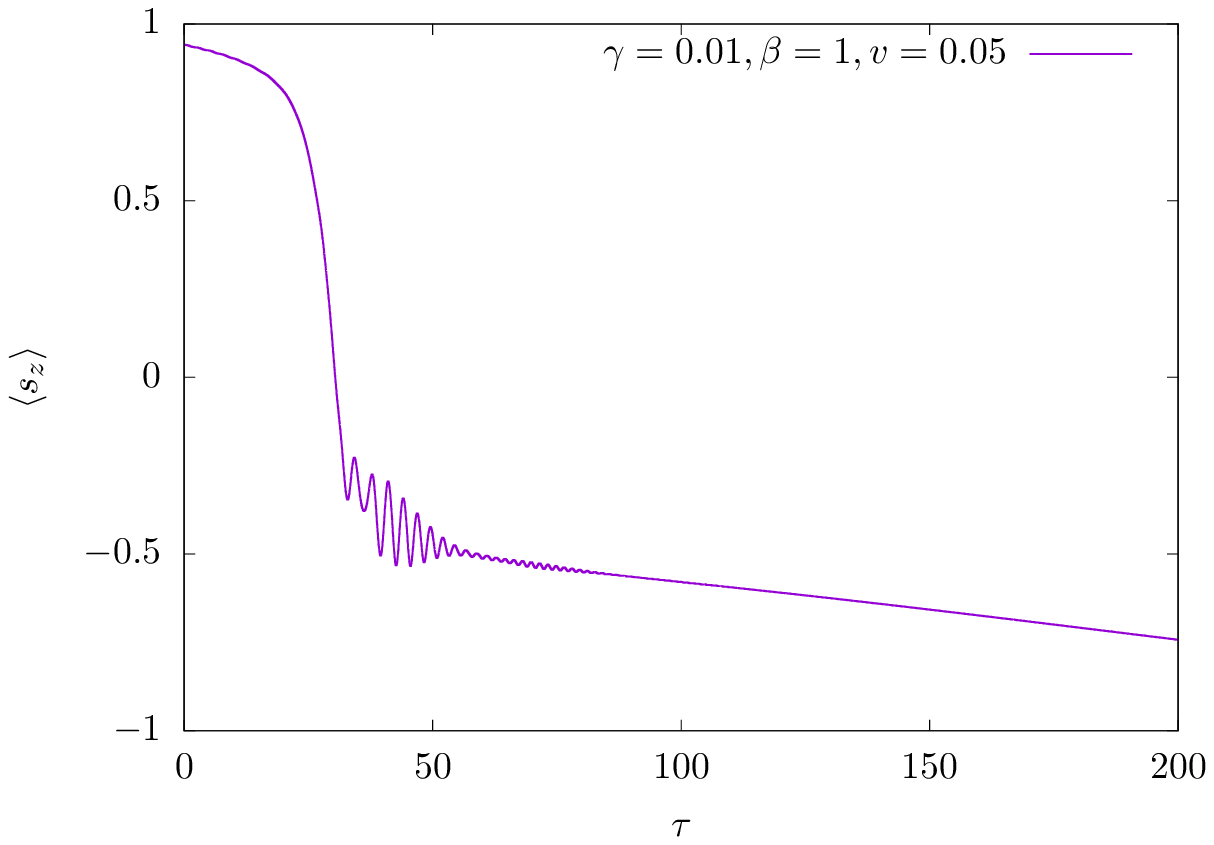} \\
\includegraphics[width=8cm]{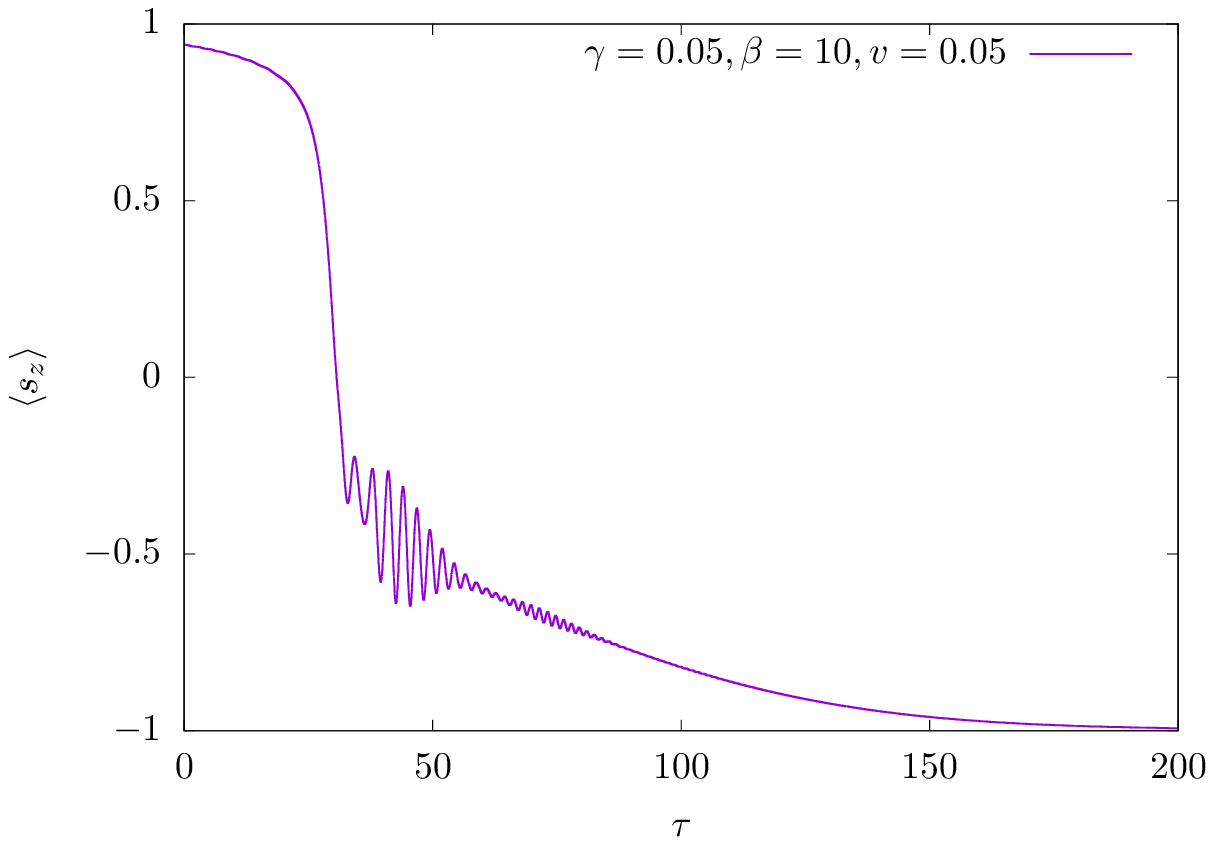} & \includegraphics[width=8cm]{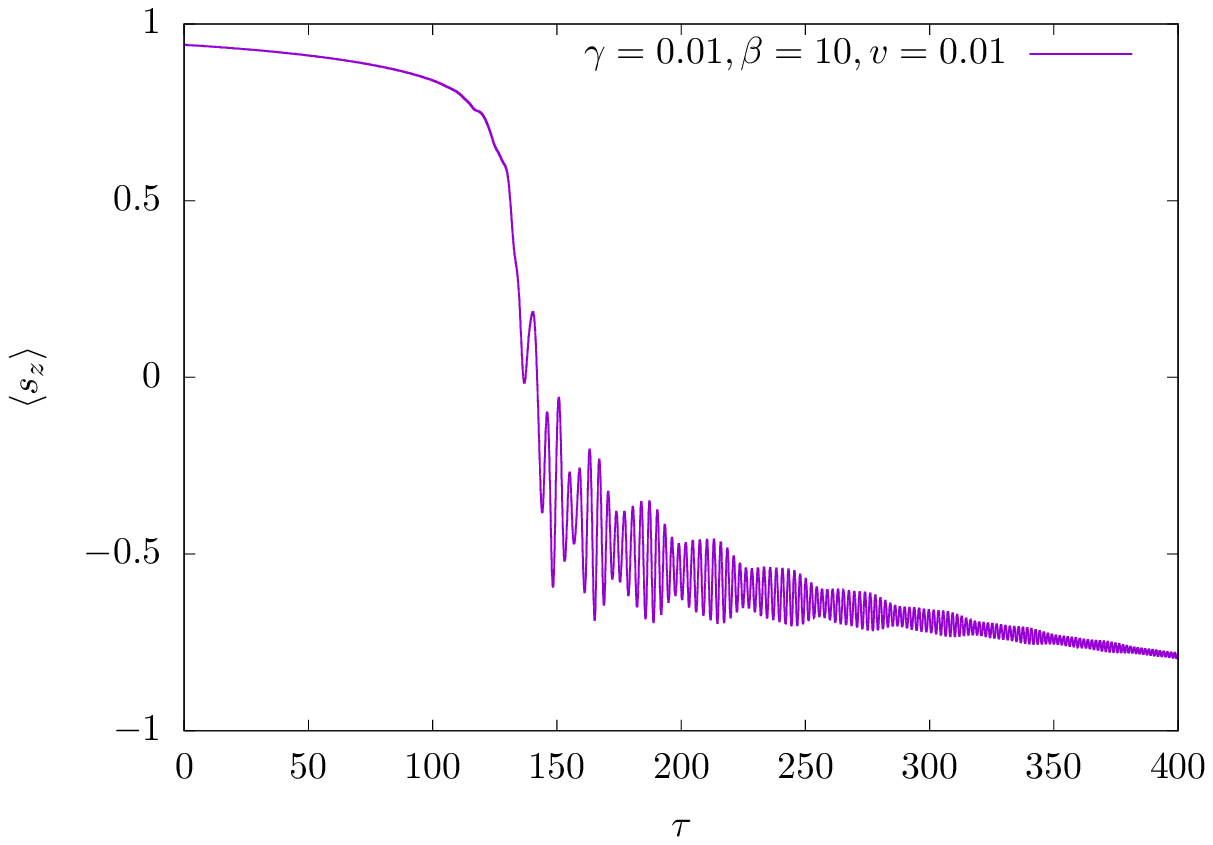} 
\end{tabular}
\caption{\label{dis}Beatings in dissipative environments. Spin-size is $S=10$, the anisotropic constant is $D=1$, and the transverse field is $H_x=1$. The longitudinal field is swept from $H_z(0)=1$. The horizontal axis is the classical time $\tau$ and the vertical axis is the expectation value of the normalized spin operator $\langle s_z\rangle$. The cases of (top left) the coupling constant $\gamma=0.01$, the inverse temperature $\beta=10$, and the sweeping rate $v=0.05$, (top right) $\gamma=0.01$, $\beta=1$, and $v=0.05$, (bottom left) $\gamma=0.05$, $\beta=10$, $v=0.05$, and (bottom right) $\gamma=0.01$, $\beta=10$, and $v=0.01$ are plotted. }
\end{figure*}
%

The beating phenomenon of the spin-length has been found in the magnetization dynamics after the Stoner-Wohlfarth point~\cite{HBM}. 
This recursive oscillation is regarded as 
the beating of the resonant oscillations between the adjacent energy levels. 
The period of this beating oscillation is given by
\begin{equation}
T_\tau=\frac{2\pi S}{D},\quad\tilde{T}_\tau=\frac{2\pi}{\tilde{D}},
\label{period}
\end{equation}
in the $\tau$ unit. 
We remark that this beating is not only the oscillation of the $z$ component of magnetization~\cite{GS}, but also of all the components of spin, and as the result, of the spin length
\begin{equation}
s_f=\langle s_x\rangle^2+\langle s_y\rangle^2+\langle s_z\rangle^2,
\label{sfid}
\end{equation}
which we call the spin-fidelity~\cite{HBM} (Fig.~\ref{beatfig}). 

Surprisingly, it was also found that the period of the recursive oscillation does not depend on the strength of the magnetic field $H_z$. 
In this sense, the beating is robust. 
However, in order to observe the beating in experiments at finite temperatures, we must study their stability against relaxation and decoherence.

\subsection{Effects of relaxation and decoherence}

In order to take dissipative effects in account, we use the generalized Lindblad-type equation~\cite{STM}. 
Although the general treatment for the time-dependent fields is difficult~\cite{Shirai}, the generalized Lindblad equation works in the cases, where the change of parameters is much slower than the relaxation time of the thermal bath~\cite{STM,Weidlich1965,KTH,Shirai,MM}. 

In this scheme, we consider the quantum Stoner-Wohlfarth model coupled with a thermal reservoir. 
We assume that the interaction Hamiltonian and the bath Hamiltonian are given by
\begin{eqnarray}
\mathcal{H}_I&=&\lambda s_x(B+B^\dag),\quad B=\sum_kg_kb_k, \\
\mathcal{H}_B&=&\sum_k\omega_kb_k^\dag b_k,
\end{eqnarray}
where $\lambda$ is the coupling constant, $b_k$ is the bosonic operator associated with the thermal barth, and $\omega_k$ and $g_k$ are the characteristic parameters of the thermal reservoir. 
With such an interaction Hamiltonian, both spin dephasing and relaxation are taken in account. 

In the weak coupling limit $\lambda\ll1$, the dissipative dynamics is given by the generalized Lindblad equation~\cite{STM,MM}
\begin{equation}
\frac{\partial}{\partial t}\rho(t)=-i[\mathcal{H}(t),\rho(t)]-\lambda^2\pi\{[s_x,R(t)\rho(t)]+\mathrm{h.c.}\},
\label{genLind}
\end{equation}
with 
\begin{eqnarray}
R(t)&=&\sum_{k,l}\Psi(E_k(t)-E_l(t)) \nonumber \\
&&\quad\times\langle\psi_k(t)|s_x|\psi_l(t)\rangle|\psi_k(t)\rangle\langle\psi_l(t)|,
\end{eqnarray}
where $\rho(t)$ is the reduced density operator for the system, $E_k(t)$ is the instantaneous eigenenergy of the quantum Stoner-Wohlfarth model
\begin{equation}
\mathcal{H}(t)|\psi_k(t)\rangle=E_k(t)|\psi_k(t)\rangle,
\end{equation}
and $\Psi(\cdot)$ is the bath spectral function 
\begin{equation}
\Psi(\omega)=\frac{J(\omega)-J(-\omega)}{e^{\beta\omega}-1},\quad J(\omega)=AS\omega^l\Theta(\omega). 
\end{equation}
Here, $\beta$ is the inverse temperature, $J(\omega)$ is the spectral density, and $\Theta(\omega)$ is the step function. 
Spin-size $S$ appears in the spectral density $J(\omega)$ due to the normalization of the system Hamiltonian $\mathcal{H}(t)$. 
The thermal reservoir is called Ohmic for $l=1$, sub-Ohmic for $l<1$, and super-Ohmic for $l>1$. 
In the following, we calculate the beating dynamics in the Ohmic case $l=1$ for different values of the (inverse) temperature $\beta$, the coupling-constant $\gamma=\lambda^2\pi AS$, and the sweeping-rate $v$ (Fig.~\ref{dis}). 

The oscillations of magnetizations with beatings, observed in the previous paper~\cite{HBM}, exhibit two time scales as seen in Fig.~\ref{beatfig}. 
The faster one is the simple spin-precession with the time-dependent frequency $\tilde{\omega}_p\sim2\tilde{D}\langle S_z\rangle+\tilde{H}_z$ and the slower one is the beating with the time and field-strength independent frequency $\tilde{\omega}_b\sim\tilde{D}$. 
Detection of both frequencies, i.e. of the full spin motion, should be possible in the absence of damping if the measurement frequency $\omega_m$ is faster than both frequencies, $\omega_m\gg\omega_p>\omega_b$. 
However, this is difficult to realize in real systems. 
Furthermore, the spin dynamics is generally damped by the environments. 
This results in finite spin-lattice (dissipation) and spin-spin (decoherence) times $T_1$  and $T_2$, respectively. 
In the Ohmic generalized Lindblad-type equation, the parameter is $\gamma\propto1/T_1$. 
In Fig.~\ref{dis}, we show the effects of the inverse temperature $\beta$ and the coupling constant $\gamma$. 
On the experimental side, measurements may require ensembles of identical single-domain ferromagnetic nano-particles, single-molecular magnets, or atomic magnets. 
Of course, single-objects measurements are also possible. 

Taking the example of the single-molecular magnet $\mathrm{Mn}_{12}$, the experiment~\cite{TLBGSB} shows that the anisotropic constant $\tilde{D}=0.61\mathrm{K}$ giving the ratio $\tilde{D}/g\mu_B=0.44\mathrm{T}$. 
For an infinitesimal transverse field $H_x$, the Stoner-Wohlfarth point for $\mathrm{Mn}_{12}$ is given by $H_\mathrm{SW}\simeq8.8\mathrm{T}$. 
The frequency of the beating is given by $\omega_b\simeq1.3\times10\mathrm{GHz}$. 
In spite of the fact that such a value is rather large, we believe that these beatings could be observed in a particular setup which will be described later. 
Furthermore, other systems such as single-spin magnets should show significantly smaller beating frequencies. 

%
%
\section{\label{sum}Summary and discussions}
In the present paper, we extended our previous work on the quantum Stoner-Wohlfarth model~\cite{HBM} (\ref{QSWham}) to the studies of (i) the distributions of the eigenstate populations and their associated scaling properties (Sec.~\ref{distribution}) and (ii) the beating dynamics of magnetization in the dissipative (thermal) environment (Sec.~\ref{dissipative}). 

(i) At a given swept field, the distribution of the eigenstate populations beyond the Stoner-Wohlfarth point is given by the amount of the scattered populations at the successive avoided level crossings along the metastable branch, which is the diabatic continuation of the state corresponding to $M_z=+S$ to the negative field region until the Stoner-Wohlfarth point. 
The calculations show that the distribution of eigenstate populations $\{P_k\}$ has a peak at the field $H_\mathrm{peak}$, which is not located at the Stoner-Wohlfarth point $H_\mathrm{SW}$ but is shifted a certain amount $\Delta H_\mathrm{peak}=|H_\mathrm{peak}-H_\mathrm{SW}|$, which depends on the sweeping rate $v$. 
It should be noted that the shift does not depend on $S$, and thus it is the same in the classical limit $S\to\infty$. 
We investigated the $S$ dependence of the distribution of the eigenstate populations $\{P_k\}$, and found a possible scaling form. 
The dependence of the shift on $v$ was estimated to be proportional to $v^{2/3}$ for small $v$, which was discussed from the viewpoint of the scattering at each avoided crossing and the associated criticality around the Stoner-Wohlfarth point. 

(ii) Finally, we studied how the beatings found outside of the Stoner-Wohlfarth point $H_z< H_\mathrm{SW}<0$ are modified by dissipative effects due to the contact with a thermal reservoir. 
Adopting the generalized Lindblad-type equation~\cite{STM}, we showed how the beatings could be preserved while magnetization relaxes to the ground state, and clarified how a fast enough measurement and sweeping time-scales could allow their observation at low enough temperatures. 

%
%
\begin{acknowledgments}
This work is supported by Grants-in-Aid for Scientific Research C (Grant No. 25400391) from MEXT of Japan and the Elements Strategy Initiative Center for Magnetic Materials under the outsourcing project of MEXT. 
The numerical calculations were supported by the supercomputer center of ISSP of the University of Tokyo. 
T. H. is supported by the Program for Leading Graduate Schools: Material Education program for the future leaders in Research, Industry, and Technology (MERIT) from JSPS. 
\end{acknowledgments}
%
%
%

\bibliography{qsw_bib}

\begin{thebibliography}{31}%
\makeatletter
\providecommand \@ifxundefined [1]{%
 \@ifx{#1\undefined}
}%
\providecommand \@ifnum [1]{%
 \ifnum #1\expandafter \@firstoftwo
 \else \expandafter \@secondoftwo
 \fi
}%
\providecommand \@ifx [1]{%
 \ifx #1\expandafter \@firstoftwo
 \else \expandafter \@secondoftwo
 \fi
}%
\providecommand \natexlab [1]{#1}%
\providecommand \enquote  [1]{``#1''}%
\providecommand \bibnamefont  [1]{#1}%
\providecommand \bibfnamefont [1]{#1}%
\providecommand \citenamefont [1]{#1}%
\providecommand \href@noop [0]{\@secondoftwo}%
\providecommand \href [0]{\begingroup \@sanitize@url \@href}%
\providecommand \@href[1]{\@@startlink{#1}\@@href}%
\providecommand \@@href[1]{\endgroup#1\@@endlink}%
\providecommand \@sanitize@url [0]{\catcode `\\12\catcode `\$12\catcode
  `\&12\catcode `\#12\catcode `\^12\catcode `\_12\catcode `\%12\relax}%
\providecommand \@@startlink[1]{}%
\providecommand \@@endlink[0]{}%
\providecommand \url  [0]{\begingroup\@sanitize@url \@url }%
\providecommand \@url [1]{\endgroup\@href {#1}{\urlprefix }}%
\providecommand \urlprefix  [0]{URL }%
\providecommand \Eprint [0]{\href }%
\providecommand \doibase [0]{http://dx.doi.org/}%
\providecommand \selectlanguage [0]{\@gobble}%
\providecommand \bibinfo  [0]{\@secondoftwo}%
\providecommand \bibfield  [0]{\@secondoftwo}%
\providecommand \translation [1]{[#1]}%
\providecommand \BibitemOpen [0]{}%
\providecommand \bibitemStop [0]{}%
\providecommand \bibitemNoStop [0]{.\EOS\space}%
\providecommand \EOS [0]{\spacefactor3000\relax}%
\providecommand \BibitemShut  [1]{\csname bibitem#1\endcsname}%
\let\auto@bib@innerbib\@empty
\bibitem [{\citenamefont {N\'eel}(1947)}]{Neel}%
  \BibitemOpen
  \bibfield  {author} {\bibinfo {author} {\bibfnamefont {L.}~\bibnamefont
  {N\'eel}},\ }\href@noop {} {\bibfield  {journal} {\bibinfo  {journal} {Compl.
  Rend. Acad. Sci.}\ }\textbf {\bibinfo {volume} {224}},\ \bibinfo {pages}
  {1488} (\bibinfo {year} {1947})}\BibitemShut {NoStop}%
\bibitem [{\citenamefont {Wernsdorfer}\ \emph
  {et~al.}(1997{\natexlab{a}})\citenamefont {Wernsdorfer}, \citenamefont
  {Bonet~Orozco}, \citenamefont {Hasselbach}, \citenamefont {Benoit},
  \citenamefont {Mailly}, \citenamefont {Kubo}, \citenamefont {Nakano},\ and\
  \citenamefont {Barbara}}]{PhysRevLett.79.4014}%
  \BibitemOpen
  \bibfield  {author} {\bibinfo {author} {\bibfnamefont {W.}~\bibnamefont
  {Wernsdorfer}}, \bibinfo {author} {\bibfnamefont {E.}~\bibnamefont
  {Bonet~Orozco}}, \bibinfo {author} {\bibfnamefont {K.}~\bibnamefont
  {Hasselbach}}, \bibinfo {author} {\bibfnamefont {A.}~\bibnamefont {Benoit}},
  \bibinfo {author} {\bibfnamefont {D.}~\bibnamefont {Mailly}}, \bibinfo
  {author} {\bibfnamefont {O.}~\bibnamefont {Kubo}}, \bibinfo {author}
  {\bibfnamefont {H.}~\bibnamefont {Nakano}}, \ and\ \bibinfo {author}
  {\bibfnamefont {B.}~\bibnamefont {Barbara}},\ }\href {\doibase
  10.1103/PhysRevLett.79.4014} {\bibfield  {journal} {\bibinfo  {journal}
  {Phys. Rev. Lett.}\ }\textbf {\bibinfo {volume} {79}},\ \bibinfo {pages}
  {4014} (\bibinfo {year} {1997}{\natexlab{a}})}\BibitemShut {NoStop}%
\bibitem [{\citenamefont {Sessoli}\ \emph {et~al.}(1993)\citenamefont
  {Sessoli}, \citenamefont {Gatteschi}, \citenamefont {Caneschi},\ and\
  \citenamefont {Novak}}]{SGCN}%
  \BibitemOpen
  \bibfield  {author} {\bibinfo {author} {\bibfnamefont {R.}~\bibnamefont
  {Sessoli}}, \bibinfo {author} {\bibfnamefont {D.}~\bibnamefont {Gatteschi}},
  \bibinfo {author} {\bibfnamefont {A.}~\bibnamefont {Caneschi}}, \ and\
  \bibinfo {author} {\bibfnamefont {M.~A.}\ \bibnamefont {Novak}},\ }\href@noop
  {} {\bibfield  {journal} {\bibinfo  {journal} {Nature}\ }\textbf {\bibinfo
  {volume} {365}},\ \bibinfo {pages} {141} (\bibinfo {year}
  {1993})}\BibitemShut {NoStop}%
\bibitem [{\citenamefont {Gatteschi}\ \emph {et~al.}(2006)\citenamefont
  {Gatteschi}, \citenamefont {Sessoli},\ and\ \citenamefont {Villain}}]{GSV}%
  \BibitemOpen
  \bibfield  {author} {\bibinfo {author} {\bibfnamefont {D.}~\bibnamefont
  {Gatteschi}}, \bibinfo {author} {\bibfnamefont {R.}~\bibnamefont {Sessoli}},
  \ and\ \bibinfo {author} {\bibfnamefont {J.}~\bibnamefont {Villain}},\
  }\href@noop {} {\emph {\bibinfo {title} {Molecular Nanomagnets}}}\ (\bibinfo
  {publisher} {Oxford University Press},\ \bibinfo {year} {2006})\BibitemShut
  {NoStop}%
\bibitem [{\citenamefont {Giraud}\ \emph {et~al.}(2001)\citenamefont {Giraud},
  \citenamefont {Wernsdorfer}, \citenamefont {Tkachuk}, \citenamefont
  {Mailly},\ and\ \citenamefont {Barbara}}]{PhysRevLett.87.057203}%
  \BibitemOpen
  \bibfield  {author} {\bibinfo {author} {\bibfnamefont {R.}~\bibnamefont
  {Giraud}}, \bibinfo {author} {\bibfnamefont {W.}~\bibnamefont {Wernsdorfer}},
  \bibinfo {author} {\bibfnamefont {A.~M.}\ \bibnamefont {Tkachuk}}, \bibinfo
  {author} {\bibfnamefont {D.}~\bibnamefont {Mailly}}, \ and\ \bibinfo {author}
  {\bibfnamefont {B.}~\bibnamefont {Barbara}},\ }\href {\doibase
  10.1103/PhysRevLett.87.057203} {\bibfield  {journal} {\bibinfo  {journal}
  {Phys. Rev. Lett.}\ }\textbf {\bibinfo {volume} {87}},\ \bibinfo {pages}
  {057203} (\bibinfo {year} {2001})}\BibitemShut {NoStop}%
\bibitem [{\citenamefont {Giraud}\ \emph {et~al.}(2003)\citenamefont {Giraud},
  \citenamefont {Tkachuk},\ and\ \citenamefont
  {Barbara}}]{PhysRevLett.91.257204}%
  \BibitemOpen
  \bibfield  {author} {\bibinfo {author} {\bibfnamefont {R.}~\bibnamefont
  {Giraud}}, \bibinfo {author} {\bibfnamefont {A.~M.}\ \bibnamefont {Tkachuk}},
  \ and\ \bibinfo {author} {\bibfnamefont {B.}~\bibnamefont {Barbara}},\ }\href
  {\doibase 10.1103/PhysRevLett.91.257204} {\bibfield  {journal} {\bibinfo
  {journal} {Phys. Rev. Lett.}\ }\textbf {\bibinfo {volume} {91}},\ \bibinfo
  {pages} {257204} (\bibinfo {year} {2003})}\BibitemShut {NoStop}%
\bibitem [{\citenamefont {Thomas}\ \emph {et~al.}(1996)\citenamefont {Thomas},
  \citenamefont {Lionti}, \citenamefont {Ballou}, \citenamefont {Gatteschi},
  \citenamefont {Sessoli},\ and\ \citenamefont {Barbara}}]{TLBGSB}%
  \BibitemOpen
  \bibfield  {author} {\bibinfo {author} {\bibfnamefont {L.}~\bibnamefont
  {Thomas}}, \bibinfo {author} {\bibfnamefont {F.}~\bibnamefont {Lionti}},
  \bibinfo {author} {\bibfnamefont {R.}~\bibnamefont {Ballou}}, \bibinfo
  {author} {\bibfnamefont {D.}~\bibnamefont {Gatteschi}}, \bibinfo {author}
  {\bibfnamefont {R.}~\bibnamefont {Sessoli}}, \ and\ \bibinfo {author}
  {\bibfnamefont {B.}~\bibnamefont {Barbara}},\ }\href@noop {} {\bibfield
  {journal} {\bibinfo  {journal} {Nature}\ }\textbf {\bibinfo {volume} {383}},\
  \bibinfo {pages} {145} (\bibinfo {year} {1996})}\BibitemShut {NoStop}%
\bibitem [{\citenamefont {Barbara}(2012)}]{B}%
  \BibitemOpen
  \bibfield  {author} {\bibinfo {author} {\bibfnamefont {B.}~\bibnamefont
  {Barbara}},\ }\href@noop {} {\bibfield  {journal} {\bibinfo  {journal} {Phil.
  Trans. R. Soc. A}\ }\textbf {\bibinfo {volume} {370}},\ \bibinfo {pages}
  {4487} (\bibinfo {year} {2012})}\BibitemShut {NoStop}%
\bibitem [{\citenamefont {Landau}(1932)}]{L}%
  \BibitemOpen
  \bibfield  {author} {\bibinfo {author} {\bibfnamefont {L.~D.}\ \bibnamefont
  {Landau}},\ }\href@noop {} {\bibfield  {journal} {\bibinfo  {journal} {Phys.
  Z. Sowjetunion}\ }\textbf {\bibinfo {volume} {2}},\ \bibinfo {pages} {46}
  (\bibinfo {year} {1932})}\BibitemShut {NoStop}%
\bibitem [{\citenamefont {Zener}(1932)}]{Z}%
  \BibitemOpen
  \bibfield  {author} {\bibinfo {author} {\bibfnamefont {C.}~\bibnamefont
  {Zener}},\ }\href@noop {} {\bibfield  {journal} {\bibinfo  {journal} {Proc.
  R. Soc. London Ser. A}\ }\textbf {\bibinfo {volume} {137}},\ \bibinfo {pages}
  {696} (\bibinfo {year} {1932})}\BibitemShut {NoStop}%
\bibitem [{\citenamefont {Majonara}(1932)}]{M}%
  \BibitemOpen
  \bibfield  {author} {\bibinfo {author} {\bibfnamefont {E.}~\bibnamefont
  {Majonara}},\ }\href@noop {} {\bibfield  {journal} {\bibinfo  {journal}
  {Nuovo Cimento}\ }\textbf {\bibinfo {volume} {9}},\ \bibinfo {pages} {43}
  (\bibinfo {year} {1932})}\BibitemShut {NoStop}%
\bibitem [{\citenamefont {St{\"u}ckelberg}(1932)}]{S}%
  \BibitemOpen
  \bibfield  {author} {\bibinfo {author} {\bibfnamefont {E.~C.~G.}\
  \bibnamefont {St{\"u}ckelberg}},\ }\href@noop {} {\bibfield  {journal}
  {\bibinfo  {journal} {Helv. Phys. Acta}\ }\textbf {\bibinfo {volume} {5}},\
  \bibinfo {pages} {369} (\bibinfo {year} {1932})}\BibitemShut {NoStop}%
\bibitem [{\citenamefont {Miyashita}(1995)}]{doi:10.1143/JPSJ.64.3207}%
  \BibitemOpen
  \bibfield  {author} {\bibinfo {author} {\bibfnamefont {S.}~\bibnamefont
  {Miyashita}},\ }\href {\doibase 10.1143/JPSJ.64.3207} {\bibfield  {journal}
  {\bibinfo  {journal} {J. Phys. Soc. Jpn.}\ }\textbf {\bibinfo {volume}
  {64}},\ \bibinfo {pages} {3207} (\bibinfo {year} {1995})}\BibitemShut
  {NoStop}%
\bibitem [{\citenamefont {Miyashita}(1996)}]{doi:10.1143/JPSJ.65.2734}%
  \BibitemOpen
  \bibfield  {author} {\bibinfo {author} {\bibfnamefont {S.}~\bibnamefont
  {Miyashita}},\ }\href {\doibase 10.1143/JPSJ.65.2734} {\bibfield  {journal}
  {\bibinfo  {journal} {J. Phys. Soc. Jpn.}\ }\textbf {\bibinfo {volume}
  {65}},\ \bibinfo {pages} {2734} (\bibinfo {year} {1996})}\BibitemShut
  {NoStop}%
\bibitem [{\citenamefont {De~Raedt}\ \emph {et~al.}(1997)\citenamefont
  {De~Raedt}, \citenamefont {Miyashita}, \citenamefont {Saito}, \citenamefont
  {Garc\'{\i}a-Pablos},\ and\ \citenamefont {Garc\'{\i}a}}]{PhysRevB.56.11761}%
  \BibitemOpen
  \bibfield  {author} {\bibinfo {author} {\bibfnamefont {H.}~\bibnamefont
  {De~Raedt}}, \bibinfo {author} {\bibfnamefont {S.}~\bibnamefont {Miyashita}},
  \bibinfo {author} {\bibfnamefont {K.}~\bibnamefont {Saito}}, \bibinfo
  {author} {\bibfnamefont {D.}~\bibnamefont {Garc\'{\i}a-Pablos}}, \ and\
  \bibinfo {author} {\bibfnamefont {N.}~\bibnamefont {Garc\'{\i}a}},\ }\href
  {\doibase 10.1103/PhysRevB.56.11761} {\bibfield  {journal} {\bibinfo
  {journal} {Phys. Rev. B}\ }\textbf {\bibinfo {volume} {56}},\ \bibinfo
  {pages} {11761} (\bibinfo {year} {1997})}\BibitemShut {NoStop}%
\bibitem [{\citenamefont {Wernsdorfer}\ and\ \citenamefont
  {Sessoli}(1999)}]{WS}%
  \BibitemOpen
  \bibfield  {author} {\bibinfo {author} {\bibfnamefont {W.}~\bibnamefont
  {Wernsdorfer}}\ and\ \bibinfo {author} {\bibfnamefont {R.}~\bibnamefont
  {Sessoli}},\ }\href@noop {} {\bibfield  {journal} {\bibinfo  {journal}
  {Science}\ }\textbf {\bibinfo {volume} {284}},\ \bibinfo {pages} {133}
  (\bibinfo {year} {1999})}\BibitemShut {NoStop}%
\bibitem [{\citenamefont {Ueda}\ \emph {et~al.}(2002)\citenamefont {Ueda},
  \citenamefont {Maegawa},\ and\ \citenamefont {Kitagawa}}]{UMK}%
  \BibitemOpen
  \bibfield  {author} {\bibinfo {author} {\bibfnamefont {M.}~\bibnamefont
  {Ueda}}, \bibinfo {author} {\bibfnamefont {S.}~\bibnamefont {Maegawa}}, \
  and\ \bibinfo {author} {\bibfnamefont {S.}~\bibnamefont {Kitagawa}},\
  }\href@noop {} {\bibfield  {journal} {\bibinfo  {journal} {Phys. Rev. B}\
  }\textbf {\bibinfo {volume} {66}},\ \bibinfo {pages} {073309} (\bibinfo
  {year} {2002})}\BibitemShut {NoStop}%
\bibitem [{\citenamefont {Owerre}\ and\ \citenamefont
  {Paranjape}(2015)}]{Owerre2015}%
  \BibitemOpen
  \bibfield  {author} {\bibinfo {author} {\bibfnamefont {S.~A.}\ \bibnamefont
  {Owerre}}\ and\ \bibinfo {author} {\bibfnamefont {M.~B.}\ \bibnamefont
  {Paranjape}},\ }\href@noop {} {\bibfield  {journal} {\bibinfo  {journal}
  {Phys. Rep.}\ }\textbf {\bibinfo {volume} {546}},\ \bibinfo {pages} {1}
  (\bibinfo {year} {2015})}\BibitemShut {NoStop}%
\bibitem [{\citenamefont {Stoner}\ and\ \citenamefont {Wohlfarth}(1948)}]{SW}%
  \BibitemOpen
  \bibfield  {author} {\bibinfo {author} {\bibfnamefont {E.~C.}\ \bibnamefont
  {Stoner}}\ and\ \bibinfo {author} {\bibfnamefont {E.~P.}\ \bibnamefont
  {Wohlfarth}},\ }\href@noop {} {\bibfield  {journal} {\bibinfo  {journal}
  {Phil. Trans. R. Soc. A}\ }\textbf {\bibinfo {volume} {240}},\ \bibinfo
  {pages} {599} (\bibinfo {year} {1948})}\BibitemShut {NoStop}%
\bibitem [{\citenamefont {Hatomura}\ \emph {et~al.}(2016)\citenamefont
  {Hatomura}, \citenamefont {Barbara},\ and\ \citenamefont {Miyashita}}]{HBM}%
  \BibitemOpen
  \bibfield  {author} {\bibinfo {author} {\bibfnamefont {T.}~\bibnamefont
  {Hatomura}}, \bibinfo {author} {\bibfnamefont {B.}~\bibnamefont {Barbara}}, \
  and\ \bibinfo {author} {\bibfnamefont {S.}~\bibnamefont {Miyashita}},\
  }\href@noop {} {\bibfield  {journal} {\bibinfo  {journal} {Phys. Rev. Lett.}\
  }\textbf {\bibinfo {volume} {116}},\ \bibinfo {pages} {037203} (\bibinfo
  {year} {2016})}\BibitemShut {NoStop}%
\bibitem [{\citenamefont {Nielsen}\ and\ \citenamefont {Chuang}(2000)}]{NC}%
  \BibitemOpen
  \bibfield  {author} {\bibinfo {author} {\bibfnamefont {M.~A.}\ \bibnamefont
  {Nielsen}}\ and\ \bibinfo {author} {\bibfnamefont {I.~L.}\ \bibnamefont
  {Chuang}},\ }\href@noop {} {\emph {\bibinfo {title} {Quantum Computation and
  Quantum Information}}}\ (\bibinfo  {publisher} {Cambridge University Press},\
  \bibinfo {year} {2000})\BibitemShut {NoStop}%
\bibitem [{\citenamefont {Breuer}\ and\ \citenamefont
  {Petruccione}(2002)}]{BP}%
  \BibitemOpen
  \bibfield  {author} {\bibinfo {author} {\bibfnamefont {H.~P.}\ \bibnamefont
  {Breuer}}\ and\ \bibinfo {author} {\bibfnamefont {F.}~\bibnamefont
  {Petruccione}},\ }\href@noop {} {\emph {\bibinfo {title} {The Theory of Open
  Quantum Systems}}}\ (\bibinfo  {publisher} {Oxford University Press},\
  \bibinfo {year} {2002})\BibitemShut {NoStop}%
\bibitem [{\citenamefont {Weiss}(2012)}]{W}%
  \BibitemOpen
  \bibfield  {author} {\bibinfo {author} {\bibfnamefont {U.}~\bibnamefont
  {Weiss}},\ }\href@noop {} {\emph {\bibinfo {title} {Quantum Dissipative
  Systems}}},\ \bibinfo {edition} {4th}\ ed.\ (\bibinfo  {publisher} {World
  Scientific},\ \bibinfo {year} {2012})\BibitemShut {NoStop}%
\bibitem [{\citenamefont {Saito}\ \emph {et~al.}(2000)\citenamefont {Saito},
  \citenamefont {Takesue},\ and\ \citenamefont {Miyashtia}}]{STM}%
  \BibitemOpen
  \bibfield  {author} {\bibinfo {author} {\bibfnamefont {K.}~\bibnamefont
  {Saito}}, \bibinfo {author} {\bibfnamefont {S.}~\bibnamefont {Takesue}}, \
  and\ \bibinfo {author} {\bibfnamefont {S.}~\bibnamefont {Miyashtia}},\
  }\href@noop {} {\bibfield  {journal} {\bibinfo  {journal} {Phys. Rev. E}\
  }\textbf {\bibinfo {volume} {61}},\ \bibinfo {pages} {2397} (\bibinfo {year}
  {2000})}\BibitemShut {NoStop}%
\bibitem [{\citenamefont {Wernsdorfer}\ \emph
  {et~al.}(1997{\natexlab{b}})\citenamefont {Wernsdorfer}, \citenamefont
  {Orozco}, \citenamefont {Barbara}, \citenamefont {Hasselbach}, \citenamefont
  {Benoit}, \citenamefont {Mailly}, \citenamefont {Doudin}, \citenamefont
  {Meier}, \citenamefont {Wegrowe}, \citenamefont {Ansermet}, \citenamefont
  {Demoncy}, \citenamefont {Pascard}, \citenamefont {Demoncy}, \citenamefont
  {Loiseau}, \citenamefont {Francois}, \citenamefont {Duxin},\ and\
  \citenamefont {Pileni}}]{doi:10.1063/1.364656}%
  \BibitemOpen
  \bibfield  {author} {\bibinfo {author} {\bibfnamefont {W.}~\bibnamefont
  {Wernsdorfer}}, \bibinfo {author} {\bibfnamefont {E.~B.}\ \bibnamefont
  {Orozco}}, \bibinfo {author} {\bibfnamefont {B.}~\bibnamefont {Barbara}},
  \bibinfo {author} {\bibfnamefont {K.}~\bibnamefont {Hasselbach}}, \bibinfo
  {author} {\bibfnamefont {A.}~\bibnamefont {Benoit}}, \bibinfo {author}
  {\bibfnamefont {D.}~\bibnamefont {Mailly}}, \bibinfo {author} {\bibfnamefont
  {B.}~\bibnamefont {Doudin}}, \bibinfo {author} {\bibfnamefont
  {J.}~\bibnamefont {Meier}}, \bibinfo {author} {\bibfnamefont {J.~E.}\
  \bibnamefont {Wegrowe}}, \bibinfo {author} {\bibfnamefont {J.-P.}\
  \bibnamefont {Ansermet}}, \bibinfo {author} {\bibfnamefont {N.}~\bibnamefont
  {Demoncy}}, \bibinfo {author} {\bibfnamefont {H.}~\bibnamefont {Pascard}},
  \bibinfo {author} {\bibfnamefont {N.}~\bibnamefont {Demoncy}}, \bibinfo
  {author} {\bibfnamefont {A.}~\bibnamefont {Loiseau}}, \bibinfo {author}
  {\bibfnamefont {L.}~\bibnamefont {Francois}}, \bibinfo {author}
  {\bibfnamefont {N.}~\bibnamefont {Duxin}}, \ and\ \bibinfo {author}
  {\bibfnamefont {M.~P.}\ \bibnamefont {Pileni}},\ }\href {\doibase
  10.1063/1.364656} {\bibfield  {journal} {\bibinfo  {journal} {J. Appl.
  Phys.}\ }\textbf {\bibinfo {volume} {81}},\ \bibinfo {pages} {5543} (\bibinfo
  {year} {1997}{\natexlab{b}})}\BibitemShut {NoStop}%
\bibitem [{\citenamefont {Mori}\ \emph {et~al.}(2010)\citenamefont {Mori},
  \citenamefont {Miyashita},\ and\ \citenamefont
  {Rikvold}}]{PhysRevE.81.011135}%
  \BibitemOpen
  \bibfield  {author} {\bibinfo {author} {\bibfnamefont {T.}~\bibnamefont
  {Mori}}, \bibinfo {author} {\bibfnamefont {S.}~\bibnamefont {Miyashita}}, \
  and\ \bibinfo {author} {\bibfnamefont {P.~A.}\ \bibnamefont {Rikvold}},\
  }\href {\doibase 10.1103/PhysRevE.81.011135} {\bibfield  {journal} {\bibinfo
  {journal} {Phys. Rev. E}\ }\textbf {\bibinfo {volume} {81}},\ \bibinfo
  {pages} {011135} (\bibinfo {year} {2010})}\BibitemShut {NoStop}%
\bibitem [{\citenamefont {Garanin}\ and\ \citenamefont {Schilling}(2004)}]{GS}%
  \BibitemOpen
  \bibfield  {author} {\bibinfo {author} {\bibfnamefont {D.~A.}\ \bibnamefont
  {Garanin}}\ and\ \bibinfo {author} {\bibfnamefont {R.}~\bibnamefont
  {Schilling}},\ }\href@noop {} {\bibfield  {journal} {\bibinfo  {journal}
  {Phys. Rev. B}\ }\textbf {\bibinfo {volume} {69}},\ \bibinfo {pages} {104412}
  (\bibinfo {year} {2004})}\BibitemShut {NoStop}%
\bibitem [{\citenamefont {Shirai}\ \emph {et~al.}(2014)\citenamefont {Shirai},
  \citenamefont {Mori},\ and\ \citenamefont {Miyashita}}]{Shirai}%
  \BibitemOpen
  \bibfield  {author} {\bibinfo {author} {\bibfnamefont {T.}~\bibnamefont
  {Shirai}}, \bibinfo {author} {\bibfnamefont {T.}~\bibnamefont {Mori}}, \ and\
  \bibinfo {author} {\bibfnamefont {S.}~\bibnamefont {Miyashita}},\ }\href@noop
  {} {\bibfield  {journal} {\bibinfo  {journal} {J. Phys. B: At. Mol. Opt.
  Phys.}\ }\textbf {\bibinfo {volume} {47}},\ \bibinfo {pages} {025501}
  (\bibinfo {year} {2014})}\BibitemShut {NoStop}%
\bibitem [{\citenamefont {Weidlich}\ and\ \citenamefont
  {Haake}(1965)}]{Weidlich1965}%
  \BibitemOpen
  \bibfield  {author} {\bibinfo {author} {\bibfnamefont {W.}~\bibnamefont
  {Weidlich}}\ and\ \bibinfo {author} {\bibfnamefont {F.}~\bibnamefont
  {Haake}},\ }\href@noop {} {\bibfield  {journal} {\bibinfo  {journal} {Z.
  Phys.}\ }\textbf {\bibinfo {volume} {185}},\ \bibinfo {pages} {30} (\bibinfo
  {year} {1965})}\BibitemShut {NoStop}%
\bibitem [{\citenamefont {Kubo}\ \emph {et~al.}(1991)\citenamefont {Kubo},
  \citenamefont {Toda},\ and\ \citenamefont {Hashitsume}}]{KTH}%
  \BibitemOpen
  \bibfield  {author} {\bibinfo {author} {\bibfnamefont {R.}~\bibnamefont
  {Kubo}}, \bibinfo {author} {\bibfnamefont {M.}~\bibnamefont {Toda}}, \ and\
  \bibinfo {author} {\bibfnamefont {N.}~\bibnamefont {Hashitsume}},\
  }\href@noop {} {\emph {\bibinfo {title} {Statical Physics II}}}\ (\bibinfo
  {publisher} {Springer-Verlag Berlin Heidelberg},\ \bibinfo {year}
  {1991})\BibitemShut {NoStop}%
\bibitem [{\citenamefont {Mori}\ and\ \citenamefont {Miyashita}(2008)}]{MM}%
  \BibitemOpen
  \bibfield  {author} {\bibinfo {author} {\bibfnamefont {T.}~\bibnamefont
  {Mori}}\ and\ \bibinfo {author} {\bibfnamefont {S.}~\bibnamefont
  {Miyashita}},\ }\href@noop {} {\bibfield  {journal} {\bibinfo  {journal} {J.
  Phys. Soc. Jpn.}\ }\textbf {\bibinfo {volume} {77}},\ \bibinfo {pages}
  {124005} (\bibinfo {year} {2008})}\BibitemShut {NoStop}%
\end{thebibliography}%

\end{document}